\documentclass[preprint,aps,showpacs,prd,tightenlines]{revtex4}
\usepackage{graphicx}
\usepackage[latin1]{inputenc}
\newcommand{\be}{\begin{eqnarray}} 
\newcommand{\ee}{\end{eqnarray}}
\newcommand{\nn}{~\nonumber \\}
\newcommand{\p}{{\cal P}\exp}

\newcommand{\ssh}{\gamma\cdot}

\newcommand{\bmp}{\noindent\begin{minipage}{16cm}}
\newcommand{\emp}{\end{minipage}\vskip 7mm} 
\newcommand{\tilda}{\check}

\begin{document}


\title{
Quark-antiquark pair production in space-time dependent fields
}

\author{Dennis D. Dietrich}
\affiliation{
Laboratoire de Physique Th\'eorique, Universit\'e Paris XI, Orsay, France}

\date{October 4, 2004}


\begin{abstract}

Fermion-antifermion pair-production in the presence of classical fields is
described based on the retarded and advanced fermion propagators. They are
obtained by solving the equation of motion for the Dirac Green's functions
with the respective boundary conditions to all orders in the field.
Subsequently, various approximation schemes fit for different field
configurations are explained. This includes longitudinally boost-invariant
forms. Those occur frequently in the description of ultrarelativistic
heavy-ion collisions in the semiclassical limit. As a next step, the gauge
invariance of the expression for the expectation value of the number of
produced fermion-antifermion pairs as a functional of said propagators is
investigated in detail. Finally, the calculations are carried out for a
longitudinally boost-invariant model-field, taking care of the last issue,
especially.

\end{abstract}



\maketitle


\section{Introduction}

In \cite{field} the inclusive spectra of fermion-antifermion pairs produced
by vacuum polarisation in the presence of time-dependent classical fields
were studied based on the exact retarded propagator. Here the generalisation
to arbitrarily space-time dependent fields is presented. Processes of this kind in
quantum electrodynamics (QED) \cite{baltz,qed} as well as in quantum 
chromodynamics (QCD) \cite{qcd} are
of importance for the physics of the early universe \cite{cornwall} and
strong laser fields \cite{laser} as well as of ultrarelativistic heavy-ion
collisions and the quark-gluon plasma (QGP) \cite{kinetic}.
The strong classical fields are the common feature of these phenomena, while 
the detailed characteristics of the field configurations differ. Although
this paper ultimately concentrates on
ultrarelativistic heavy-ion collisions, various other kinematical regimes
are treated as well (see section \ref{theretprop}). 

A lot of effort is made to study the QGP's production and equilibration
\cite{qm} in nuclear collision experiments at the Relativistic
Heavy-Ion Collider (RHIC) and the Large Hadron Collider (LHC), currently
under construction. 
It is a widely used hypothesis that the initial state in heavy-ion collisions 
is dominated by gluons which on account of their large occupation number can be 
treated as a classical background field.
The larger the occupation number $N_k$ of the bosonic sector at a given 
momentum $k$ of a physical system, {\it i.e.}, the better \mbox{$N_k\gg 1$}
is satisfied, the more accurately it can be described by a classical field. 
In a heavy-ion collision at RHIC with $\sqrt{s}=130$GeV the initial
occupation number for gluons at their average transverse momentum
$|\vec k_T|\approx 1$GeV in the center of the collision is approximately 
equal to 1.5 \cite{mueller}. 
Even if this number is not much larger than unity, the classical field as
the expectation value of the gauge field still represents the  
parametrically leading contribution.
At this scale, keeping only classical bosons is a first 
approximation but quantum fluctuations have to be investigated subsequently.

One concrete model based on this idea is due to McLerran and
Venugopalan \cite{jimwlk}. In that approach, classically interacting 
colour-charge 
distributions on the two branches of the light-cone represent the two 
colliding nuclei. However, there, it turns out that quantum corrections 
become important due to kinematical factors. 
Nevertheless, this approach can be extended farther by including quantum 
evolution in the distribution of the classically interacting charges. This
extension leads to the so-called JIMWLK equation \cite{jimwlk}. 

Be that as it may,
the magnitude of the high occupation-number bosonic fields is such that
multiple couplings to the classical field, {\it i.e.}, higher powers of
the product of the coupling constant $g$ and the gauge field strength $A$,
are not parametrically suppressed.
In the absence of other scales they have to be addressed to all orders.
Under the prerequisite of weak coupling the leading quantum processes
concern terms in the classical action of second order in the quantum fields
(fermions, antifermions, and bosonic quantum-fluctuations).
These terms' coefficients give the inverse of the respective two-point 
Green's functions. Their inversion for selected boundary conditions yields the 
propagators of the corresponding particles to all orders in the classical
field.
As an alternative, the full propagator can also be derived by resumming all
terms of the perturbative series or by adding up a complete set of
wave-function solutions of the equations of motion.

Since exact solutions are not easily obtainable some general approximation
schemes have been devised.
Expansion of the propagators in powers of the field leads to the
perturbative series which is based on the free Green's functions.
Neglecting the spatial part of the field and the derivatives yields what is
called the static approximation.
Other approaches for the fermionic propagator can be found in 
\cite{gromes}.

In this paper the phenomenon of particle production is to be investigated
based on the propagator in the classical field.
The required correlator is obtained by direct solution of the equation of
motion.
In QED as well as QCD the direct production of fermion-antifermion
pairs occurs (electron-positron and quark-antiquark, respectively). 
Caused by the non-linearity of the gauge-field tensor in QCD also pairs of
gluonic quantum fluctuations are created.
There, in strong fields the two types of reactions are equally
parametrically favoured.

In a given situation the bosonic sector could be covered sufficiently well by 
the concept of a classical field and
the hard sector be accessible perturbatively \cite{ddd}.
However, due to Pauli's principle, it exists no such concept for the
fermions.
Perturbative investigations can only describe the hard not the soft
part.
Hence, this paper deals predominantly with the description 
of fermions and antifermions. 

In constant fields, the Schwinger mechanism \cite{schwinger} gives the
exact answer to the problem of particle production. Frequently, its
application is extended to slowly varying fields and/or low-energy particles. 
It, like most of the other schemes, is based on assumptions on
time and/or energy scales of the situation under investigation. However,
if such scales are to be obtained by means of a calculation it is
likely that they are strongly biased by the choice of the approximation
\cite{comparison}.

Finally, it is necessary to decide whether expectation values are to be
obtained in the framework of the {\it in-in}-formalism of quantum field 
theory -- {\it i.e.}, based on the advanced and retarded propagator and their
relatives -- or probabilities within that of the {\it in-out}-formalism --
{\it i.e.}, starting out from the Feynman propagator -- \cite{field,baltz}.
As the average number of produced pairs is the more natural observable as
compared to the probability to produce exactly a given number $n$ of pairs
in a particular event, the first possibility will be pursued here. 

Therefore, section \ref{theretprop} presents the solution of the
equation of motion for the retarded Dirac propagator 
to all orders in the field. Respective subsections include details on a
generalised translation operator that occurs during the derivation as well
as on the propagator in a large variety of expansion schemes and classes of
field configurations. This includes longitudinally
boost-invariant field configurations which can be of importance in the
description of the initial phase of an ultrarelativistic heavy-ion collisions.
Section III contains the application of the results of section II to the 
problem of particle production in general. This includes a thorough
discussion of the gauge invariance of the result. In a subsection
the formulae are evaluated for a longitudinally boost-invariant model-field.
In the last chapter the contents of the paper are summarised.

Throughout the paper the metric tensor is given by: $g^{\mu\nu}={\rm
diag}(1,-1,-1,-1)$, angular momenta are measured in units of $\hbar$, and
velocities in fractions of the speed of light $c$. From hereon, the coupling
constant is included in the definition of the classical field: 
\mbox{$gA^\mu_{old}=A^\mu_{new}$}. The convention for Fourier
transformations of one-point functions is:

\be
f(x)
=
\int
\frac{d^4k}{(2\pi)^4}
e^{-ik\cdot x}
f(k),
\ee

\noindent
that for two-point functions:

\be
f(x,y)
=
\int
\frac{d^4p}{(2\pi)^4}
\frac{d^4q}{(2\pi)^4}
e^{-iq\cdot x}
e^{+ip\cdot y}
f(q,p).
\ee


\section{The retarded fermion-propagator\label{theretprop}}

\begin{center}
\begin{figure}[h]
\resizebox{!}{3cm}{\includegraphics{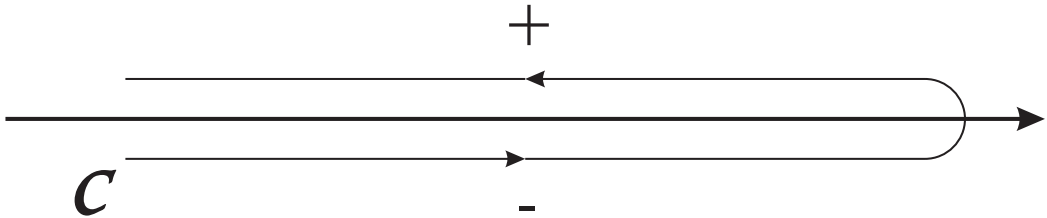}}
\caption{Contour of integration in the complex time-plane for the 
Schwinger-Keldysch formalism}
\label{skcontour}
\end{figure}
\end{center}

If the fermionic propagators for the {\it in-in}-formalism are known to all
orders in the classical field the generating functional of the Green's
functions without radiative corrections can be written as:

\be
Z[{\bf\bar\eta},{\bf\eta}]
=
\exp
\left\{
i\int d^4xd^4y~
\bar\eta_{\sigma_1}(x)G_{\sigma_1\sigma_2}(x,y)\eta_{\sigma_2}(y)
\right\},
\ee

\noindent
with implicit sums over $\sigma_{1,2}\in\{+,-\}$. These indices refer to the sign of
the imaginary part of the time component of $x$ or $y$ respectively, taken along the
contour in the complex time-plane shown in figure \ref{skcontour} \cite{sk}. 
Hence, the components of the matrix propagator $G_{\sigma_1\sigma_2}$ 
correspond to:

\bmp
\be
G_{++}(x,y)&=&+\left<0_{in}\right|T\bar\psi(y)\psi(x)\left|0_{in}\right>
\nn
G_{--}(x,y)&=&+\left<0_{in}\right|T^{-1}\bar\psi(y)\psi(x)\left|0_{in}\right>
\nn
G_{+-}(x,y)&=&+\left<0_{in}\right|\bar{\psi}(y)\psi(x)\left|0_{in}\right>
\nn
G_{-+}(x,y)&=&-\left<0_{in}\right|\psi(x)\bar\psi(y)\left|0_{in}\right>,
\ee
\emp

\noindent
where $T^{-1}$ stands for reverse time-ordering.
The components of the matrix propagator $G_{\sigma_1\sigma_2}$ can be linked to the
advanced (A), retarded (R), and on-shell (S) propagator by:

\be
2G_{\sigma_1\sigma_2}
=
G_S+\sigma_1 G_A+\sigma_2 G_R.
\label{trafo}
\ee

\noindent
What is called the on-shell propagator is a solution of the homogeneous
Dirac equation. The advanced and the retarded propagator solve the
equation of motion of the Dirac Green's functions:

\be
\left[
i\ssh\partial(x)+\ssh A(x)-m
\right]
G(x,y)
=
\delta^{(4)}(x-y).
\ee

\noindent
Let us consider homogeneous solutions \mbox{$G_H(x,y)$} of Dirac's equation,
{\it i.e.}, with the $\delta$-distribution replaced by zero using a product 
as an ansatz:

\be
G_H(x,y)=U(x_0-y_0)\hat G_H(x,y).
\label{prodans}
\ee

\noindent
The differential equation obeyed by the functional $U(x_0-y_0)$ is to
be the following:

\be
[i\gamma^0\partial_0(x)U(x_0-y_0)]\hat G_H(x,y)
+
i\gamma^j\partial_j(x)U(x_0-y_0)\hat G_H(x,y)
=
0,
\label{diffeqtransop}
\ee

\noindent
with a summation over \mbox{$j\in\{1,2,3\}$} and where the solution for the 
boundary condition \mbox{$U(0)=1$} is given by:

\be
U(x_0-y_0)=\exp[-(x_0-y_0)\gamma^0\gamma^j\partial_j(x)].
\label{gentransop}
\ee

\noindent
This object corresponds to a generalised translation operator taking care of 
the relativity of events, as shall be explained in section \ref{gto}.
With equation (\ref{diffeqtransop}) and after
subsequent multiplication by $-iU(y_0-x_0)\gamma^0$ from the left, the
remaining differential equation for the second factor \mbox{$\hat G_H(x,y)$} 
of the homogeneous solution reads:

\be
\partial_0(x)\hat G_H(x,y)
-
iU(y_0-x_0)\gamma^0[\ssh A(x)-m]U(x_0-y_0)\hat G_H(x,y)
=
0,
\ee

\noindent
where the inverse of the generalised translation operator (\ref{gentransop}) 
is given by inverting the sign of its argument: 
\mbox{$U^{-1}(x_0-y_0)=U(y_0-x_0)$}.
For the boundary condition $G_H(x_0=y_0,\vec x,\vec y)=1$ the last
differential equation is solved
by a path-ordered exponential \cite{field}:

\be
\hat G_H(x,y)
=
\p\left\{
i\int_{y_0}^{x_0}d\xi_0
U(y_0-\xi_0)
\gamma^0
\left[
\ssh A(\xi_0,\vec x)-m
\right]
U(\xi_0-y_0)
\right\}.
\label{secfac}
\ee

\noindent
Other boundary conditions would only lead to additional factors to the right
of the above solution which would not result in independent expressions,
because a homogeneous differential equation is investigated here.
The homogeneous solution $G_H(x,y)$ can be reassembled from equations
(\ref{gentransop}) and (\ref{secfac}) according to equation (\ref{prodans}).
The boundary condition for vanishing time difference \mbox{$x_0-y_0$} for
all Dirac Green's functions is given by
\mbox{$[\lim_{x_0\rightarrow y_0+0}-\lim_{x_0\rightarrow y_0-0}]G(x,y)
=-i\gamma^0\delta^{(3)}(\vec x-\vec y)$}.
Additionally, the retarded propagator \mbox{$G_R(x,y)$} is required to vanish
for negative time differences: \mbox{$G_R(x,y)=0~{\rm if}~x_0<y_0$}; the
advanced for positive: \mbox{$G_A(x,y)=0~{\rm if}~x_0>y_0$}.
By virtue of the boundary condition chosen for equation (\ref{secfac}) the
retarded propagator can be expressed as:

\be
iG_R(x,y)\gamma^0
=
+
G_H(x,y)
\delta^{(3)}(\vec x-\vec y)
\theta(x_0-y_0),
\ee

\noindent
the advanced as:

\be
iG_A(x,y)\gamma^0
=
-
G_H(x,y)
\delta^{(3)}(\vec x-\vec y)
\theta(y_0-x_0).
\ee

With the help of the general resummation formula derived in \cite{field}:

\bmp
\be
&&\p\left\{\int^{x_0}_{y_0}d\xi_0[B(\xi_0)+C(\xi_0)]\right\}
=
\nn
&=&
\p\left\{\int^{x_0}_{y_0}d\xi_0 B(\xi_0)\right\}
\times
\nn
&&\times
\p\left[
\int^{x_0}_{y_0}d\xi_0
\p\left\{\int_{\xi_0}^{y_0}dz_0 B(z_0)\right\}
C(\xi_0)
\p\left\{\int^{\xi_0}_{y_0}dz_0 B(z_0)\right\}
\right]
=
\nn
&=&
\p\left[
\int^{x_0}_{y_0}d\xi_0
\p\left\{\int_{\xi_0}^{x_0}dz_0 B(z_0)\right\}
C(\xi_0)
\p\left\{\int^{\xi_0}_{x_0}dz_0 B(z_0)\right\}
\right]
\times
\nn
&&\times
\p\left\{\int^{x_0}_{y_0}d\xi_0 B(\xi_0)\right\},
\label{genresum}
\ee
\emp

\noindent
which is based on the group property valid for path-ordered exponentials:

\be
\p\left\{\int_{y_0}^{x_0}d\xi_0B(\xi_0)\right\}
=
\p\left\{\int_{z_0}^{x_0}d\xi_0B(\xi_0)\right\}
\times
\p\left\{\int_{y_0}^{z_0}d\xi_0B(\xi_0)\right\},
\label{group}
\ee

\noindent
the generalised translation operators can be included into the principal
path-ordered exponential: 

\be
&&iG_R(x,y)\gamma^0
=
\int
\frac{d^3k}{(2\pi)^3}
e^{+i\vec k\cdot(\vec x-\vec y)}
G_H(x,y,\vec k)
\theta(x_0-y_0)
=
\nn
&=&
\int
\frac{d^3k}{(2\pi)^3}
e^{+i\vec k\cdot(\vec x-\vec y)}
\p\left\{
i\int_{y_0}^{x_0}d\xi_0\gamma^0
\left[
i\gamma^j\partial_j(x)+\gamma^jk_j+\ssh A(\xi_0,\vec x)-m
\right]
\right\}
\theta(x_0-y_0)
\label{retprop}
\nn
\ee

\noindent
Here, the $\delta$-distribution has been replaced by its Fourier 
representation. The path-ordered exponential in equation (\ref{retprop})
corresponds to a special mixed representation of the homogeneous solution
$G_H$. Only
the spatial relative coordinates have been Fourier transformed to the
momentum space. The absolute coordinate is kept in the coordinate 
space. This becomes clearer by noticing that, because of the
$\delta$-distribution introduced by the boundary conditions, $\vec x$ in the
path-ordered exponential can always be replaced by $\vec y$ and also by 
\mbox{$(\vec x+\vec y)/2$}. Due to its mixed boundary conditions for positive 
and negative energy solutions, the Feynman propagator cannot be expressed in a
similar way. Henceforth, unless otherwise stated, $G(x,y)$ also without
the index $_R$ shall stand for the retarded propagator.


\subsection{Generalised translation operator\label{gto}}

In one spatial dimension, the object in equation (\ref{gentransop}) is linked to 
common translation operators \mbox{$\exp\{-t\partial_3\}$} through:

\be
U_3(+t)
=
\exp\{-t\gamma^0\gamma^3\partial_3\}
=
\rho^+
e^{-t\partial_3}
+
\rho^-
e^{+t\partial_3},
\label{gentransop1}
\ee

\noindent
where \mbox{$2\rho^\pm=1\pm\gamma^0\gamma^3$} are two projectors, whence
\mbox{$(\rho^\pm)^2=\rho^\pm$} and \mbox{$\rho^\pm\rho^\mp=0$}.
In three dimensions, the operator (\ref{gentransop}) transforms a
$\delta$-distribution localised at $\vec x=\vec y$ into:

\be
\exp\{-t\gamma^0\gamma^j\partial_j(x)\}\delta^{(3)}(\vec x-\vec y)
=
\frac{1}{4\pi}
\left[
\partial_t+\gamma^0\gamma^j\partial_j(x)
\right]
\frac
{
\delta(|\vec x-\vec y|-t)
-
\delta(|\vec x-\vec y|+t)
}
{|\vec x-\vec y|}.
\ee

\noindent
The effect on any other function can be investigated by a convolution of it
with the previous equation. The corresponding integral has only support 
in the distance $t$ from the point $\vec x$ along the light-cone, {\it i.e.}, on a 
sphere of radius $t$. After making use of the one-dimensional
$\delta$-distributions there remain two angular integrations $d\Phi$ which 
represent 
an average of the equidistant shifts over all directions 
$\hat{\vec{z}\hskip 0.5mm}(\Phi)$
\mbox{($|{\hat{\vec z\hskip 0.5mm}}|=1$)}:

\be
\exp\{-t\gamma^0\gamma^j\partial_j(x)\}f(\vec x)
=
\frac{1}{2}
\left[
\partial_t+\gamma^0\gamma^j\partial_j(x)
\right]
\int\frac{d\Phi}{4\pi}
t
[
f(\vec x+\hat{\vec{z}}t)
+
f(\vec x-\hat{\vec{z}}t)
],
\ee

\noindent
where $\vec x$ is the center of the coordinate system. The differential
operator on the right-hand side of the last equation serves to assign
the correct element of the Clifford algebra to every contribution of
the integral.


\subsection{Expansion schemes}

In this section, various expansion schemes for the retarded propagator are
presented. This shall serve to establish the connection to the results in
the purely time-dependent case \cite{field}. Like there they are also to
help to expose the characteristics of the full
expression.


\subsubsection{Weak-field expansion\label{weakfield}}

In order to obtain the weak-field expansion, the general resummation
formula (\ref{genresum}) can be used to transform the path-ordered
exponential in equation (\ref{retprop}) with the choice 
\mbox{$C(\xi_0)=i\gamma^0\ssh A(\xi_0)$}. After expanding the result
in powers of the gauge field, \mbox{$G_H=\sum_nG_H^{(n)}$}, and Fourier
transformation of the vector potential into a mixed representation one finds
up to the second order:

\be
G_H^{(0)}(x,y,\vec k)
=
\exp\{i\gamma^0[\gamma^jk_j-m](x_0-y_0)\}
\label{freeprop}
\ee

\be
G^{(1)}_H(x,y,\vec k)
&=&
i\int_{y_0}^{x_0}d\xi_0\int\frac{d^3\kappa}{(2\pi)^3}
e^{+i\vec\kappa\cdot\vec x}
\times
\nn
&&\times
\exp\{i\gamma^0[\gamma^j(k_j+\kappa_j)-m](x_0-\xi_0)\}
i\gamma^0\ssh A(\xi_0,\vec\kappa)
\times
\nn
&&\times
\exp\{i\gamma^0[\gamma^jk_j-m](\xi_0-y_0)\},
\ee

\be
G^{(2)}_H(x,y,\vec k)
&=&
i\int_{y_0}^{x_0}d\xi_0 i\int_{y_0}^{\xi_0}d\eta_0
\int\frac{d^3\kappa}{(2\pi)^3}\frac{d^3\lambda}{(2\pi)^3}
e^{+i(\vec\kappa+\vec\lambda)\cdot\vec x}
\times
\nn
&&\times
\exp\{i\gamma^0[\gamma^j(k_j+\kappa_j+\lambda_j)-m](x_0-\xi_0)\}
i\gamma^0\ssh A(\xi_0,\vec\kappa)
\times
\nn
&&\times
\exp\{i\gamma^0[\gamma^j(k_j+\lambda_j)-m](\xi_0-\eta_0)\}
i\gamma^0\ssh A(\eta_0,\vec\lambda)
\times
\nn
&&\times
\exp\{i\gamma^0[\gamma^jk_j-m](\eta_0-y_0)\}.
\ee

\noindent
Subsequent orders are obtained analogously. Replacing the exponential
functions in the higher orders by the lowest order expression in equation
(\ref{freeprop}), reproduces the usual form of the weak-field expansion.


\subsubsection{Strong-field expansion\label{sfe}}

The expansion in the last section has been adapted for fields with
\mbox{$A\ll\omega$}. In the opposite case \mbox{$A\gg\omega$}, for a strong
field, the redistribution of the exponent of the retarded propagator
according to equation (\ref{genresum}) must be such that
\mbox{$C(\xi_0)=i\gamma^0[\gamma^jk_j-m]$}.
Hence the retarded propagator becomes:

\bmp
\be
G_H(x,y,\vec k)
&=&
\p\left[
i\int_{y_0}^{x_0}d\xi_0
\p\left\{
i\int_{\xi_0}^{x_0}dz_0\gamma^0
[i\gamma^j\partial_j(x)+\ssh A(z_0,\vec x)]
\right\}
\right.
\times
\nn
&&\times
\left.
\gamma^0[\gamma^jk_j-m]
\p\left\{
i\int^{\xi_0}_{x_0}dz_0\gamma^0
[i\gamma^j\partial_j(x)+\ssh A(z_0,\vec x)]
\right\}
\right]
\times
\nn
&\times&
\p\left\{
i\int_{y_0}^{x_0}d\xi_0\gamma^0
[i\gamma^j\partial_j(x)+\ssh A(\xi_0,\vec x)]
\right\},
\label{strong}
\ee
\emp

\noindent
which subsequently can be expanded in powers of the term involving the
mass and the three-momentum. The secondary path-ordered exponentials can
again be decomposed into field insertions and generalised translation
operators:

\be
&&\p\left\{
i\int_{y_0}^{x_0}d\xi_0
\gamma^0[i\gamma^j\partial_j(x)+\ssh A(\xi_0,\vec x)]
\right\}
=
\nn
&=&
\p\left[
i\int_{y_0}^{x_0}d\xi_0
U(x_0-\xi_0)
\gamma^0\ssh A(\xi_0,\vec x)
U(\xi_0-x_0)
\right]
U(x_0-y_0)
\label{strongzero}
\ee

\noindent
Based on what was said about the general translation operator in section
\ref{gto}, the last expression shows how the behaviour of the retarded 
propagator depends on the values of the field at the relativistically retarded
positions (with a light-like displacement from $\vec x$).
That equation (\ref{strongzero}) corresponds really to an expansion in the 
on-shell energy $\omega$ can be seen from the fact that:
\mbox{$\{\gamma^0[\gamma^jk_j-m]\}^2=\omega^2$}. As the on-shell energy
$\omega$ equals the (asymptotic) relativistic mass, the strong-field
expansion represents one in the "inertia" as opposed to one in the
"accelerations" which is the weak-field expansion.


\subsubsection{Gradient expansion}

For gauge fields whose temporal gradients are much larger than their spatial
ones, a gradient expansion is obtained with the help of the general
resummation formula (\ref{genresum}) by choosing 
\mbox{$C=-\gamma^0\gamma^j\partial_j$}:

\be
G_H(x,y,\vec k)
&=&
\p\left[
i\int_{y_0}^{x_0}d\xi_0
\p\left\{
i\int_{\xi_0}^{x_0}dz_0\gamma^0[\gamma^jk_j-m+\ssh A(z_0,\vec x)]
\right\}
\right.
\times
\nn
&&~\times
\left.
i\gamma^0\gamma^j\partial_j(x)
\p\left\{
i\int^{\xi_0}_{x_0}dz_0\gamma^0[\gamma^jk_j-m+\ssh A(z_0,\vec x)]
\right\}
\right]
\times
\nn
&&\times
\p\left\{
i\int_{y_0}^{x_0}d\xi_0\gamma^0[\gamma^jk_j-m+\ssh A(\xi_0,\vec x)]
\right\}.
\label{gradient}
\ee

A result analogous to the most general in \cite{field} is obtained as the
lowest order term of an expansion of the above equation in the spatial
gradient. Higher orders contain powers of the spatial gradient but no
generalised translation operators. This is consistent with the assumption of
a slowly varying field: At the retarded positions it can be evaluated by 
means of a Taylor expansion.


\subsubsection{Combined strong-field and gradient expansion}

In the case, where the field is strong compared to the scale set by the
on-shell energy $\omega$ and above that has only relatively small spatial
gradients, an additional expansion of equation (\ref{strongzero}) in these 
yields:

\bmp
\be
&&\p\left\{
i\int_{y_0}^{x_0}d\xi_0
\gamma^0[i\gamma^j\partial_j(x)+\ssh A(\xi_0,\vec x)]
\right\}
=
\nn
&=&
\p\left[
i\int_{y_0}^{x_0}
\p\left\{
i\int_{\xi_0}^{x_0}dz_0
\gamma^0\ssh A(z_0,\vec x)
\right\}
i\gamma^0\gamma^j\partial_j(x)
\p\left\{
i\int_{\xi_0}^{x_0}dz_0
\gamma^0\ssh A(z_0,\vec x)
\right\}
\right]
\times
\nn
&\times&
\p\left\{
i\int_{y_0}^{x_0}d\xi_0\gamma^0
\ssh A(\xi_0,\vec x)
\right\}
\ee
\emp

\noindent
Finite orders of an expansion of this expression in the spatial gradients
inserted into equation (\ref{strong}) lead to the generalisation of the
strong-field approximation presented previously in \cite{field}.


\subsubsection{Abelian expansion}

This is the only expansion scheme not based on the resummation formula 
(\ref{genresum}). The $N$-th order Abelian approximation can be obtained
according to the following prescription: Starting out
with equation (\ref{retprop}), the group property (\ref{group}) is used in 
order to obtain a product of $N+1$ path-ordered exponentials. There
is a certain arbitrariness in choosing the intermediate points, but they
must be ordered in time (see below). Subsequently, generalised
translation operators are generated in every single factor by resumming the 
derivatives. Finally, the path ordering is neglected:

\bmp
\be
&&G_H(x,y,\vec k)
=
\nn
&=&
{\cal P}\prod_{\nu=0}^N
\exp\left\{
i\int_{\xi_{\nu+1}}^{\xi_\nu}dz_0
U(\xi_\nu-z_0)\gamma^0
\left[
\gamma^jk_j+\ssh A(z_0,\vec x)-m
\right]
U(z_0-\xi_\nu)
\right\}
U(\xi_\nu-\xi_{\nu+1}),
\nn
\ee
\emp

\noindent
with \mbox{$x_0=\xi_0>\xi_1>...>\xi_N>\xi_{N+1}=y_0$} because for the
retarded propagator \mbox{$x_0>y_0$}. ${\cal P}$ denotes that the factors
are ordered with respect to the index $\nu$ with the lowest index furthest
to the left.

For infinitely small interval widths (infinitely many factors) the
exact result is reobtained, because in that case the exponential
functions can be replaced by linear factors and hence also by path-ordered
exponentials. This again gives the exact retarded propagator. However, the
present Abelian approximation scheme converges faster than the expression
with linear factors \cite{field}. The above result can be interpreted as the
propagation of the fermions over the interval $[\xi_\nu,\xi_{\nu+1}]$ with 
their average canonical momentum. The average comes ever closer to the
actual value if the intervals become smaller. To the contrary, in the 
weak-field expansion, the particles are always propagated with their 
asymptotic kinematic momentum and in higher orders they only interact more
and more often with the field.


\subsection{Different classes of field configurations\label{1+1}}

In order to obtain further insight into the structure of the propagator, let us
investigate the case where the gauge field depends mostly on the coordinates in
the zero (time) and three (longitudinal) directions. Then the propagator is
treated most conveniently in a
mixed representation as a function of said coordinates and the transverse
momentum. With the choice \mbox{$B=-\gamma^0\gamma^3\partial_3$} the 
resummation formula (\ref{genresum}) leads to:

\bmp
\be
iG(x_0,y_0,x_3,y_3,\vec k_T)\gamma^0
=
\int\frac{dk_3}{2\pi}e^{+ik_3(x_3-y_3)}\theta(x_0-y_0)
\tilda G(x_0,y_0,x_3,k_3,\vec k_T)
U_3(x_0-y_0)
\label{mtzeroaux}
\ee
\emp

\noindent
with

\be
&&\tilda G(x_0,y_0,x_3,k_3,\vec k_T)
=
\p\left\{i\int_{y_0}^{x_0}d\xi_0
U_3(x_0-\xi_0)
\gamma^0
[\gamma^Jk_J-m+\ssh A(\xi_0,x_3)]
U_3(\xi_0-x_0)
\right\}
\nn
\label{untranslated}
\ee

\noindent
and implicit summations over the transverse coordinates \mbox{$J\in\{1,2\}$}.


\subsubsection{Gauge field with longitudinal and temporal components and
gradients}

\begin{center}
\begin{figure}[h]
\resizebox{!}{5cm}{\includegraphics{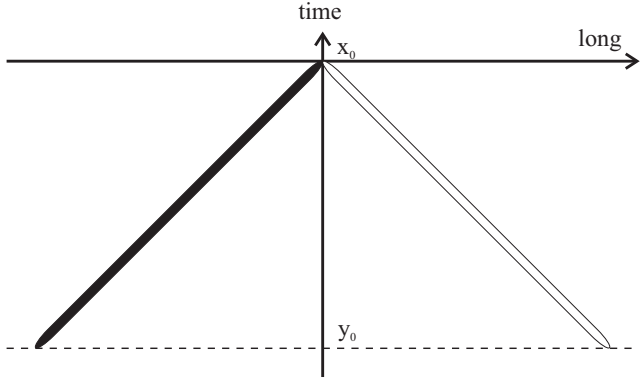}}
\caption{The two contributions to equation (\ref{mtzero}) from the backward
light-cone. The white line corresponds to the path-ordered exponential along 
$\xi_-$, the black to that along $\xi_+$. The origin of the coordinate system 
is situated at \mbox{($x_0$,$x_3$)}. The dashed line 
represents $t=y_0$.} 
\label{oneplusone}
\end{figure}
\end{center}

At first only the zero and three component of the gauge field are to be
non-vanishing. Additionally, the particles are not to have a transverse
mass. Hence, in the resulting equation
also the kernel ($C$) involving the fields and momenta can be
decomposed with the help of the projectors $\rho^\pm$ . Contributions to the 
different subspaces do not mix. Therefore, all common translation 
operators meet their inverse after acting on the components of the gauge 
field and one gets:

\be
\tilda G_{\vec A_T=\vec 0}^{m_T=0}(x_0,y_0,x_3,k_3)
&=&
\rho^+
e^{+ik_3(x_0-y_0)}
\p\left\{i\int_{\sqrt{2}(y_0-x_0)}^{0}d\xi_-
A_+(x_+,x_-+\xi_-)
\right\}
+
\nn
&+&
\rho^-
e^{-ik_3(x_0-y_0)}
\p\left\{i\int_{\sqrt{2}(y_0-x_0)}^{0}d\xi_+
A_-(x_++\xi_+,x_-)
\right\},
\label{mtzero}
\ee

\noindent
with the definition
\mbox{$
\sqrt{2}v_\pm
=
v_0\pm v_3
$}
and the change of variables to
\mbox{$
A=A(x_+,x_-)
$}.
The path-ordered exponentials on the right-hand side are Wilson lines in
which the $A_\pm$-component is integrated along the $\xi_\mp$-coordinate,
respectively. As a consequence of relativity, the endpoints of the path in
the $\xi_\pm$-direction depend on the value of the $x_\mp$-coordinate,
respectively. The two contributions which come exactly from on the backward
light-cone are schematically depicted in figure \ref{oneplusone}.

Loosening the requirements on the system by
admitting a non-zero transverse mass leads to four contributions in the
exponent of the analogous expression:

\bmp
\be
&&\tilda G_{\vec A_T=\vec 0}(x_0,y_0,x_3,k_3,\vec k_T)
~=~
\p\left[i\int_{y_0-x_0}^0d\xi_0
\right.
\times
\nn
&&~~~\times
\left(
\rho^+\left\{\sqrt{2}A_+(x_+,x_-+\sqrt{2}\xi_0)+k_3\right\}\rho^+
+
\rho^-\left\{\sqrt{2}A_-(x_++\sqrt{2}\xi_0,x_-)-k_3\right\}\rho^-
+
\right.
\nn
&&~~~+
\left.
\left.
\rho^+\gamma^0\{\gamma^Jk_J-m\}e^{+2\xi_0\partial_3(x)}\rho^-
+
\rho^-\gamma^0\{\gamma^Jk_J-m\}e^{-2\xi_0\partial_3(x)}\rho^+
\right)
\right]
\ee
\emp

\noindent
where use has been made of the idempotency of the projectors and of the
relation
\mbox{$\rho^\pm\gamma^0\{\gamma^Jk_J-m\}=\gamma^0\{\gamma^Jk_J-m\}\rho^\mp$}.
In the terms involving the transverse mass -- note that
\mbox{$[\gamma^0(\gamma^Jk_J-m)]^2={m_T}^2$} -- the translation operators do 
not meet their inverse and hence do not drop out. A resummation of this
expression for small transverse mass $m_T$ yields:

\bmp
\be
&&\tilda G_{\vec A_T=\vec 0}(x_0,y_0,x_3,k_3,\vec k_T)
~=~
\p\left[i\int_{y_0-x_0}^0d\xi_0
\tilda G_{\vec A_T=\vec 0}^{m_T=0}(x_0,x_0+\xi_0,x_3,k_3)
\right.
\times
\nn
&&~~~\times
\left(
\rho^+\gamma^0\{\gamma^Jk_J-m\}e^{+2\xi_0\partial_3(x)}\rho^-
+
\rho^-\gamma^0\{\gamma^Jk_J-m\}e^{-2\xi_0\partial_3(x)}\rho^+
\right)
\times
\nn
&&~~~\times
\left.
\tilda G_{\vec A_T=\vec 0}^{m_T=0}(x_0+\xi_0,x_0,x_3,k_3)
\right]
\tilda G_{\vec A_T=\vec 0}^{m_T=0}(x_0,y_0,x_3,k_3),
\label{smalltransversemass}
\ee
\emp

\noindent
where \mbox{$\tilda G_{\vec A_T=\vec 0}^{m_T=0}(x_0,y_0,x_3,k_3)$} has been 
taken from equation
(\ref{mtzero}). Therefore, the lowest (zero) order term in an expansion of the above
expression in the transverse mass equals the result for vanishing transverse
mass. The first order expression reads:

\bmp
\be
\tilda G^{(1)}(x_0,y_0,x_3,k_3,\vec k_T)
=
\sum_\pm
ie^{\mp ik_3(x_0-y_0)}\int_{y_0-x_0}^{0}d\xi_0e^{\mp 2i[k_3+i\partial(x)]\xi_0}
\tilda G^{(1)\pm\mp}_{\vec A_T=\vec 0},
\label{atzero}
\ee
\emp

\noindent
where the sum over the contribution with the upper signs and that with the 
lower signs is taken:

\bmp
\be
\tilda G^{(1)+-}_{\vec A_T=\vec 0}
&=&
\p\left\{i\int_{\sqrt{2}\xi_0}^{0}dz_-
A_+(x_+,x_-+z_-)
\right\}
\rho^+
\gamma^0\{\gamma^Jk_J-m\}
\rho^-
\times
\nn
&&\times
\p\left\{i\int_{\sqrt{2}(y_0-x_0-\xi_0)}^{0}dz_+
A_-(x_++z_+,x_-+\sqrt{2}\xi_0)
\right\},
\nn
\tilda G^{(1)-+}_{\vec A_T=\vec 0}
&=&
\p\left\{i\int_{\sqrt{2}\xi_0}^{0}dz_+
A_-(x_++z_+,x_-)
\right\}
\rho^-
\gamma^0\{\gamma^Jk_J-m\}
\rho^+
\times
\nn
&&\times
\p\left\{i\int_{\sqrt{2}(y_0-x_0-\xi_0)}^{0}dz_-
A_+(x_++\sqrt{2}\xi_0,x_-+z_-)
\right\}.
\label{atzeros}
\ee
\emp

The second order is sketched in figure \ref{oneplusone_transversemass}. The
kinks appear where the $m_T$-terms are inserted. Now, there are not only
contributions from on the backward light-cone but also from everywhere
within. The different orders in the expansion are weighted by powers of $m_T$
according to the number of insertions and thus with the same 
weight wherever the kink is positioned. 

For large transverse mass $m_T$ the adequate expansion would be given by an
expression analogous to the weak-field case (section \ref{weakfield}) but
without transverse components of the field.


\subsubsection{Gauge fields with dominant longitudinal and temporal gradients}

\begin{center}
\begin{figure}[h]
\resizebox{!}{5cm}{\includegraphics{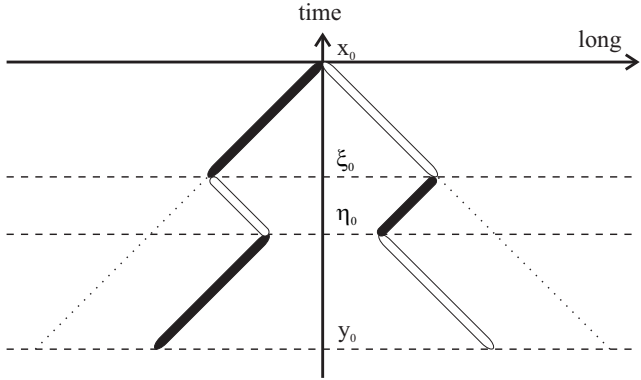}}
\caption{The two contributions to the equations (\ref{atzeros}) or
(\ref{firstorder}) respectively but for the second order.  The white lines 
correspond to path-ordered exponentials along $z_-$, the black to those along 
$z_+$. This figure also represents equation (\ref{transdom}) but where the 
black and the white lines stand for simple integration-paths. In all cases, 
the kinks are found at \mbox{$t=\xi_0$} and \mbox{$t=\eta_0$}, {\it i.e.},
where the "transverse" matrices are inserted. The origin of the coordinate 
system is situated at \mbox{($x_0$,$x_3$)}. 
The dashed lines represent lines of constant time. The dotted lines stand for 
the backward light-cone.}
\label{oneplusone_transversemass}
\end{figure}
\end{center}

Taking the transverse components of the gauge field into account, too, but 
where the dependence on time and the longitudinal components is still more
pronounced, the previous first-order expressions are modified to yield:

\noindent
\bmp
\be
\tilda G^{(1)+-}
&=&
\p\left\{i\int_{\sqrt{2}\xi_0}^{0}dz_-
A_+(x_+,x_-+z_-)
\right\}
\rho^+
{\cal M}_T(x_+,\sqrt{2}\xi_0+x_-)
\rho^-
\times
\nn
&&\times
\p\left\{i\int_{\sqrt{2}(y_0-x_0-\xi_0)}^{0}dz_+
A_-(x_++z_+,x_-+\sqrt{2}\xi_0)
\right\},
\nn
\tilda G^{(1)-+}
&=&
\p\left\{i\int_{\sqrt{2}\xi_0}^{0}dz_+
A_-(x_++z_+,x_-)
\right\}
\rho^-
{\cal M}_T(\sqrt{2}\xi_0+x_+,x_-)
\rho^+
\times
\nn
&&\times
\p\left\{i\int_{\sqrt{2}(y_0-x_0-\xi_0)}^{0}dz_-
A_+(x_++\sqrt{2}\xi_0,x_-+z_-)
\right\},
\label{firstorder}
\ee
\emp

\noindent
with 
\mbox{${\cal M_T}(x_+,x_-)=\gamma^0[\gamma^Jk_J-m+\gamma^JA_J(x_+,x_-)]$},
the contribution from the transverse vector components.
All ingredients needed in order to put together the
full expression in analogy to equation (\ref{smalltransversemass}) can
always be read off from the first-order equations, here and in the following. 
The main difference with respect to the previous section is
that now the trajectories that have a kink at a given point are no longer
weighted uniformly with \mbox{$\gamma^0\{\gamma^Jk_J-m\}$} but according to 
the value of 
\mbox{$\gamma^0\{\gamma^JA_J(x_+,x_-)+\gamma^Jk_J-m\}$} 
at the kink
\mbox{$(x_+,x_-)$} 
or, in higher order terms, the various kinks (see figure 
\ref{oneplusone_transversemass}).

In case the transverse components of the field should be dominant compared to 
the temporal and longitudinal components one can start out with equation 
(\ref{untranslated}) and resum the terms involving the transverse gauge
field:

\bmp
\be
&&\tilda G(x_0,y_0,x_3,k_3,\vec k_T)
=
\nn
&=&
\p
\left[
i\int_{y_0}^{x_0}d\xi_0
\tilda G_{A_{0,3}=0}(x_0,\xi_0,x_3,k_3,\vec k_T)
U_3(x_0-\xi_0)
\gamma^0[\gamma^3k_3+\ssh\tilde A(\xi_0,x_3)]
\right.
\times
\nn
&&\times
\left.
U_3(\xi_0-x_0)
\tilda G_{A_{0,3}=0}(\xi_0,x_0,x_3,k_3,\vec k_T)
\right]
\tilda G_{A_{0,3}=0}(x_0,y_0,x_3,k_3,\vec k_T),
\ee
\emp

\noindent
with the definition for the four-vectors without transverse components 
$\tilde v^\mu=(v^0,0,0,v^3)$ and where:

\bmp
\be
&&\tilda G_{A_{0,3}=0}(x_0,y_0,x_3,k_3,\vec k_T)
~=~
\p
\left\{
\frac{i}{\sqrt{2}}\int_{\sqrt{2}(y_0-x_0)}^0d\xi_0
\right.
\times
\nn
&&~~~\times
\left.
\left(
\rho^+
{\cal M_T}(x_+,x_-+\xi_0)
\rho^-
e^{+\sqrt{2}\xi_0\partial_3(x)}
\right.
\right.
+
\left.
\left.
\rho^-
{\cal M}_T(x_++\xi_0,x_-)
\rho^+
e^{-\sqrt{2}\xi_0\partial_3(x)}
\right)
\right\}.
\label{tildeazero}
\ee
\emp

The most important difference between the equations (\ref{tildeazero}) and
(\ref{mtzero}) is the fact, that a decomposition into
two terms involving exponentials without Clifford matrices in the argument is
not possible.
However, because of the alternating occurrence of the projectors
$\rho^\pm$, the expansion of this path-ordered exponential leads to two
contributions per order: one with $\rho^+$ on the left-hand side and the
other with a $\rho^-$. See, for example, the second-order term:
Together with the projectors, the translation operators appear in an
alternating manner. This ensures that only contributions from on or within the 
backward light-cone are taken into account. That can be seen best after putting all
translation operators to the right:

\bmp
\be
&&\tilda G^{(2)}_{A_{0,3}=0}(x_0,y_0,x_3,k_3,\vec k_T)
~=~
\frac{i}{\sqrt{2}}\int_{\sqrt{2}(y_0-x_0)}^0d\xi_0
\frac{i}{\sqrt{2}}\int_{\sqrt{2}(y_0-x_0-\xi_0)}^0d\eta_0
\times
\nn
&&~~~\times
\left[
\rho^+
{\cal M_T}(x_+,\xi_0+x_-)
\right.
{\cal M}_T(x_++\eta_0,\xi_0+x_-)
\rho^-
e^{-\sqrt{2}\eta_0\partial_3(x)}
+
\nn
&&~~~~+
\rho^-
{\cal M}_T(\xi_0+x_+,x_-)
\left.
{\cal M}_T(\xi_0+x_+,\eta_0+x_-)
\rho^+
e^{+\sqrt{2}\eta_0\partial_3(x)}
\right]
\label{transdom}
\ee
\emp

The fact that there are no contributions to the propagator from outside the
backward light-cone, consistent with relativity, also persists in the general
case (see section \ref{gto} and figure \ref{oneplusone_transversemass}).


\subsubsection{Gauge field of one time-like curvilinear coordinate}

The Green's functions obtained in \cite{field} were equal to the retarded 
propagator only for a field depending on one time-like curvilinear coordinate
\mbox{$n\cdot x$} with \mbox{$n^2>0$}.
Now one is in the position to give the retarded propagators for gauge
fields as functions of one space- \mbox{($n^2<0$)} or light-like
\mbox{($n^2=0$)} curvilinear coordinate, too. 
In order to investigate the different situations it is only necessary to look 
at one special choice for $n$ in each case, because $n^2$ is a Lorentz 
scalar. For the time-like case \mbox{$n^\mu=(1,0,0,0)$} had been chosen, for 
the space-like \mbox{$n^\mu=(0,0,0,1)$}, and for the light-like 
\mbox{$n^\mu=(1,0,0,-1)/\sqrt{2}$}. 

\begin{center}
\begin{figure}[h]
\resizebox{!}{5cm}{\includegraphics{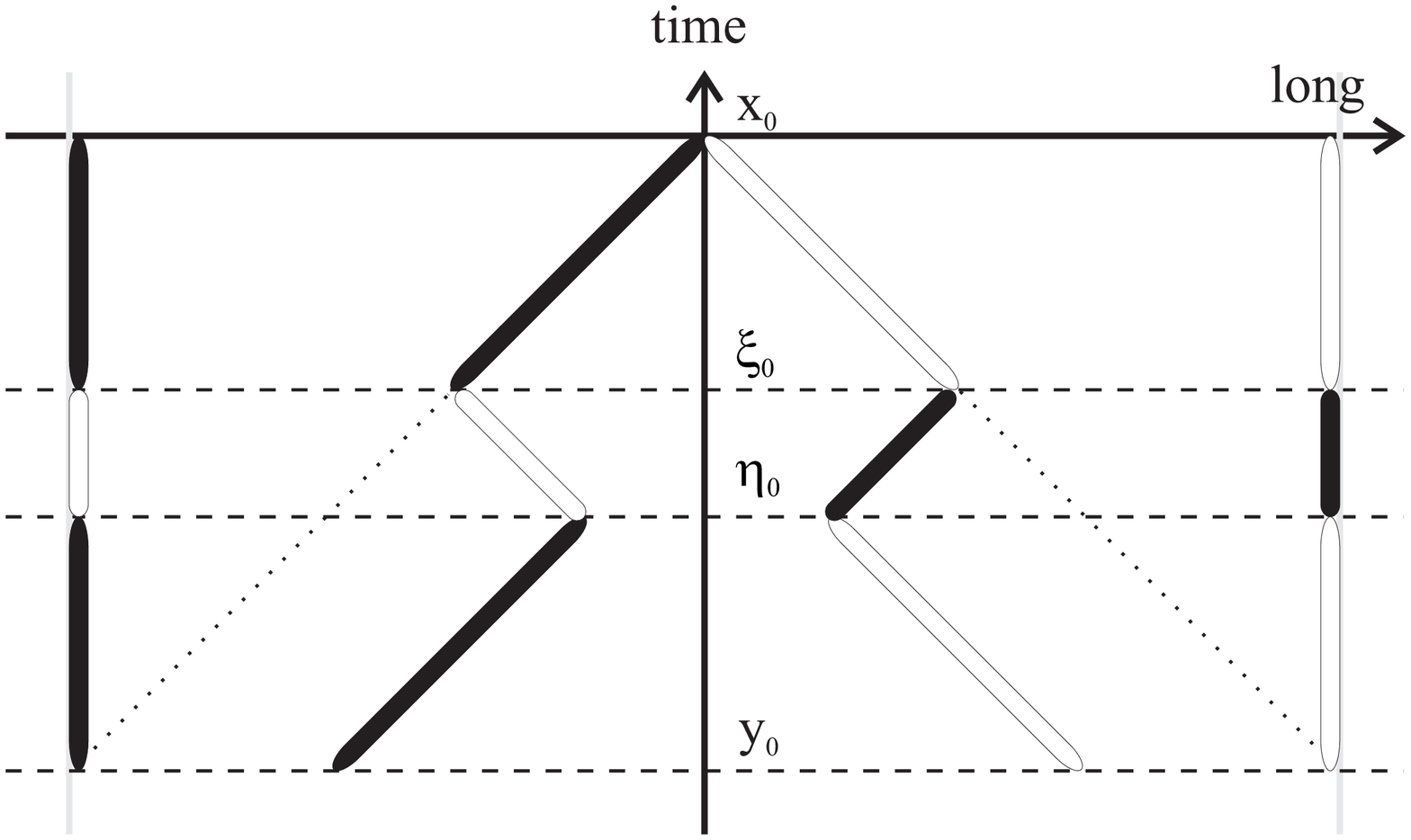}}
\caption{The two contributions in the case of a purely time-dependent
field. The sloped white lines correspond to path-ordered exponentials along
$z_-$, the black to those along $z_+$. The kinks are found at $t=\xi_0$ and
$t=\eta_0$, {\it i.e.}, where the ${\cal M}_T$ are inserted. The origin
of the coordinate system is situated at ($x_0$,$x_3$). The dashed
lines represent lines of constant time. Along these the black and white
lines in the direction of the light-cone coordinates are projected onto the
vertical ones at the sides of the figure. These projections represent the 
paths along which the path-ordered exponentials are actually evaluated in
the purely time-dependent case. The dotted lines stand for the backward
light-cone.}
\label{oneplusone_timelike}
\end{figure}
\end{center}

As already stated before, the time-like case is equivalent to the lowest-order 
expansion in the spatial gradient of equation (\ref{gradient}). A reason for
this particularly compact form can be read off figure
\ref{oneplusone_timelike}. The contributions to the propagator may be
projected along straights of constant time, {\it i.e.}, the same coordinate 
for which the retarded propagator's boundary conditions
and hence the variable for the outer integrations are given.


\subsubsection{Gauge field of one space-like curvilinear coordinate}

\begin{center}
\begin{figure}[h]
\resizebox{!}{5cm}{\includegraphics{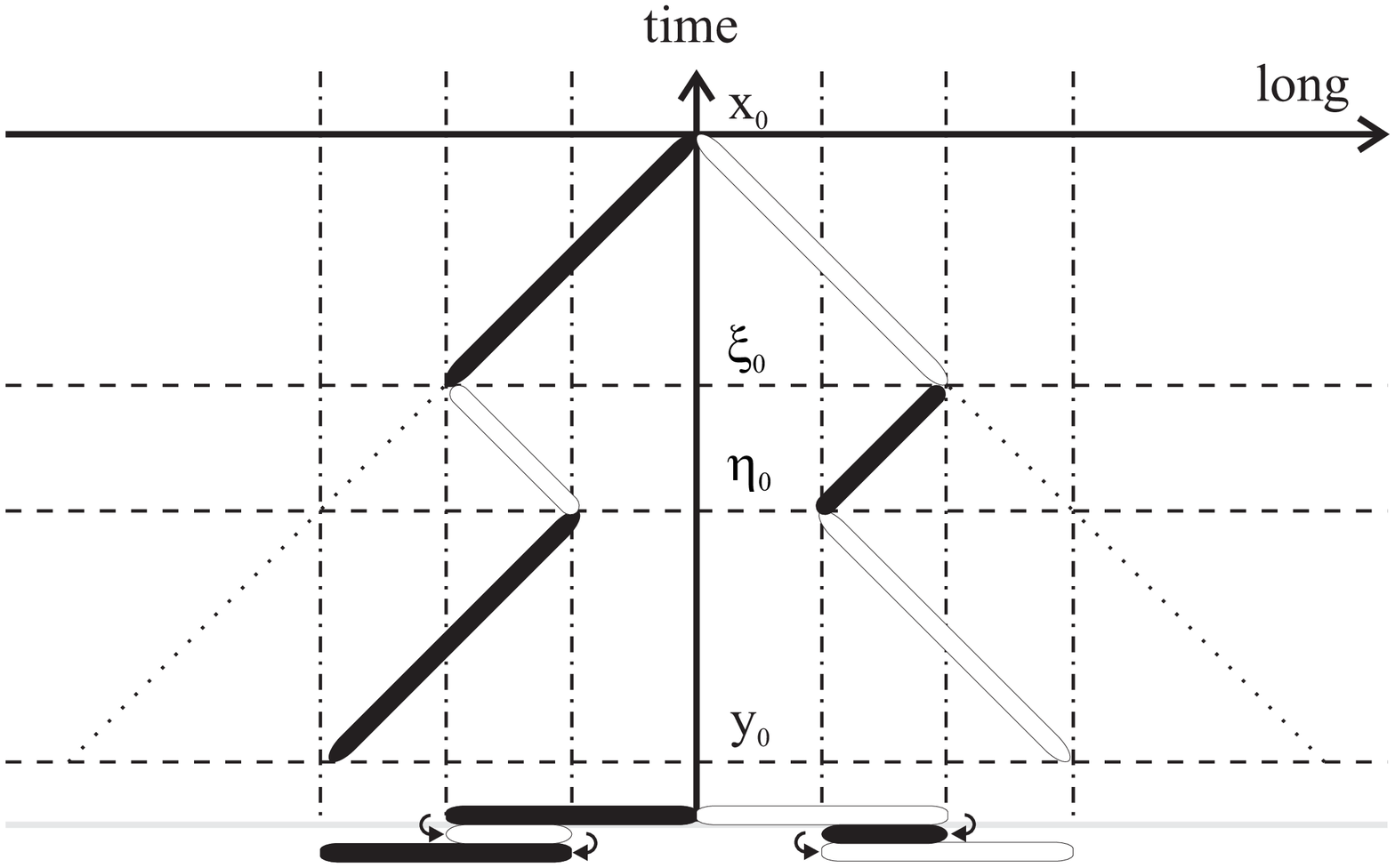}}
\vskip 5mm
\caption{The two contributions in equations (\ref{spacelike}) but for the
second order.  The sloped 
white lines correspond to path-ordered exponentials along $z_-$, the black to 
those along $z_+$. The kinks are found at \mbox{$t=\xi_0$} and 
\mbox{$t=\eta_0$}, {\it i.e.}, where the ${\cal M}_T$ are inserted. The 
origin of the coordinate system is situated at 
\mbox{($x_0$,$x_3$)}. The dashed lines represent lines of constant time. 
The dotted lines stand for the backward light-cone. The dash-dotted lines
mark straights of equal longitudinal coordinate. Along those the black and
white lines in the direction of the light-cone coordinates are projected onto the
horizontal ones at the bottom of the figure. These represent the paths along
which the path-ordered exponentials in the equations (\ref{spacelike}) are
actually evaluated.}
\label{oneplusone_spacelike}
\end{figure}
\end{center}

If the gauge field is merely a function of a space-like curvilinear
coordinate, say $x_3$, equations
(\ref{firstorder}) can be rewritten as:

\bmp
\be
&&\tilda G^{(1)+-}
=
\nn
&=&
\p\left\{i\sqrt{2}\int_{\xi_0}^{0}dz_3
A_+(x_3-z_3)
\right\}
\rho^+
{\cal M}_T(x_3-\xi_0)
\rho^-
\p\left\{i\sqrt{2}\int_{y_0-x_0}^{\xi_0}dz_3
A_-(x_3+z_3)
\right\},
\nn
&&\tilda G^{(1)-+}
=
\nn
&=&
\p\left\{i\sqrt{2}\int_{\xi_0}^{0}dz_3
A_-(x_3+z_3)
\right\}
\rho^-
{\cal M}_T(x_3+\xi_0)
\rho^+
\p\left\{i\sqrt{2}\int_{y_0-x_0}^{\xi_0}dz_3
A_+(x_3-z_3)
\right\}.
\label{spacelike}
\nn
\ee
\emp

\noindent
where the variable in the argument of the gauge field has been kept as 
\mbox{$A(x_3)$}.
As can be seen from this expression and also as indicated in figure 
\ref{oneplusone_spacelike} the field is integrated back and forth over the
longitudinal coordinate while the phase factors are still functions of time
[see equation (\ref{atzero})].
To the contrary, the respective Green's function derived for this case in
\cite{field} which is not a propagator, exhibits a different form, because a
different boundary condition had been imposed there.


\subsubsection{Gauge field of one light-like curvilinear coordinate}

For a gauge field being a function of the light-like coordinate $x_-$, the
equations (\ref{firstorder}) simplify to:

\bmp
\be
\tilda G^{(1)+-}
&=&
\p\left\{i\int_{\sqrt{2}\xi_0}^{0}dz_-
A_+(x_-+z_-)
\right\}
\rho^+
{\cal M}_T(\sqrt{2}\xi_0+x_-)
\rho^-
\times
\nn
&&\times
\exp\left\{-i\sqrt{2}(y_0-x_0-\xi_0)A_-(x_-+\sqrt{2}\xi_0)\right\},
\nn
\tilda G^{(1)-+}
&=&
\exp\left\{-i\sqrt{2}\xi_0A_-(x_-)\right\}
\rho^-
{\cal M}_T(x_-)
\rho^+
\p\left\{i\int_{\sqrt{2}(y_0-x_0-\xi_0)}^{0}dz_-
A_+(x_-+z_-)
\right\}.
\label{lightlike}
\nn
\ee
\emp

\begin{center}
\begin{figure}[h]
\resizebox{!}{5cm}{\includegraphics{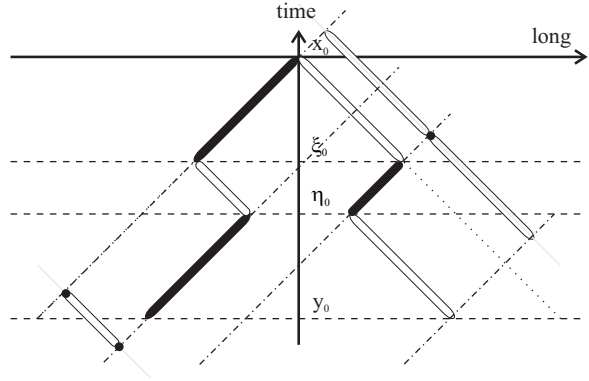}}
\vskip 5mm
\caption{The two contributions in the equations (\ref{lightlike})
but for the second order. Of the lines
on and inside the backward light-cone (dotted) the white correspond to 
path-ordered exponentials along $z_-$, the black to those along $z_+$. The 
kinks are found at \mbox{$t=\xi_0$} and \mbox{$t=\eta_0$}, {\it i.e.}, where the 
${\cal M}_T$ are inserted. The origin of the coordinate system is 
situated at \mbox{($x_0$,$x_3$)}. The dashed lines 
represent lines of constant time. The dash-dotted lines
mark straights of constant $z_-$. Along those the black and white lines
inside and on the backward light-cone are projected. These projected lines 
represent the paths along which the path-ordered exponentials in the
equations (\ref{lightlike}) are actually evaluated. 
Especially the black lines are
projected onto points which leads to simple phases as the field remains
constant.}
\label{oneplusone_lightlike}
\end{figure}
\end{center}

\noindent
Here, one of the two path-ordered exponentials per addend becomes a simple phase.
It is always the one, which is integrated along $z_+$ with $z_-$ kept
fixed. As the entire argument is only a function of $z_-$ it can be taken 
out of the integral which no longer needs path-ordering and thus only gives a 
time difference. In figure \ref{oneplusone_lightlike} these phase factors are 
the contributions parallel to the
direction of projection. Therefore, they are each mapped onto a point.
The solution presented above is of the form relevant in a field
theory with standard quantisation. The respective expression for the
homogeneous solution of the Dirac equation in the case of retarded boundary
conditions along the light cone was given in \cite{field}.


\subsubsection{Longitudinally boost-invariant gauge fields\label{lbi}}

A case of particular interest in ultrarelativistic heavy-ion collisions is
that of a gauge field which is invariant under longitudinal Lorentz
transformations, although this requirement often imposes
further limitations on the configurations of the gauge field. Independent of
this reasoning, usually also a dependence
on the transverse coordinates is present. Here the simpler case is
sufficient to
clarify the main features. As in the forward light-cone
\mbox{$\tau=\sqrt{t^2-z^2}>0$}, it does not lead to ambiguities to
replace it by half its square \mbox{$x_+x_-$}: \mbox{$A_\pm=x_\pm
a_\pm(x_+x_-)$} and \mbox{$\vec A_T=\vec A_T(x_+x_-)$}.
Therefore the equations (\ref{firstorder}) become:

\bmp
\be
\tilda G^{(1)+-}
&=&
\p\left\{i\int_{x_+(x_-+\sqrt{2}\xi_0)}^{x_+x_-}d\zeta a_+(\zeta)\right\}
\rho^+
{\cal M}_T[x_+(x_-+\sqrt{2}\xi_0)]
\rho^-
\times
\nn
&&\times
\p\left\{
i\int^{x_+(x_-+\sqrt{2}\xi_0)}_{[\sqrt{2}(y_0-\xi_0)-x_+](x_-+\sqrt{2}\xi_0)}
d\zeta a_-(\zeta)\right\},
\nn
\tilda G^{(1)-+}
&=&
\p\left\{i\int_{(x_++\sqrt{2}\xi_0)x_-}^{x_+x_-}d\zeta a_+(\zeta)\right\}
\rho^-
{\cal M}_T[(x_++\sqrt{2}\xi_0)x_-]
\rho^+
\times
\nn
&&\times
\p\left\{
i\int^{(x_++\sqrt{2}\xi_0)x_-}_{(x_++\sqrt{2}\xi_0)[\sqrt{2}(y_0-\xi_0)-x_+]}
d\zeta a_+(\zeta)\right\}
\label{boostinvariant}
\ee
\emp

\begin{center}
\begin{figure}[h]
\resizebox{!}{5cm}{\includegraphics{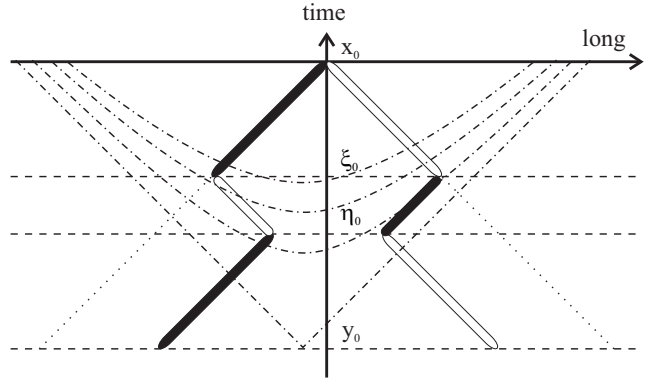}}
\vskip 5mm
\caption{The two contributions in the equations (\ref{boostinvariant}) but for
the second order. The white lines correspond to path-ordered exponentials
along $z_-$, the black to those along $z_+$. The kinks are found at 
\mbox{$t=\xi_0$} and \mbox{$t=\eta_0$}, {\it i.e.}, where the ${\cal M}_T$  
are inserted. The origin of the coordinate system is situated at 
\mbox{($x_0$,$x_3$)}. The dashed lines represent lines of constant 
time. The dotted lines stand for the backward light-cone. The dash-dotted 
hyperbolas are characterised by constant proper time $\tau$ and hence
constant values of the gauge field $A$. Mark that in general there can be an 
offset between
the origin of the coordinate system for these curves \mbox{(0,0)} and that in 
which the propagator is obtained \mbox{($x_0$,$x_3$)}. Therefore, the 
propagator does not exhibit the symmetry properties of the field.}
\label{oneplusone_boostinvariant}
\end{figure}
\end{center}

\noindent
It can be seen from these expressions that the longitudinal boost-invariance
of the vector potential does not lead to great simplifications. In fact the
propagator itself does not become boost-invariant. The reason for this is
that the boundary condition for the retarded propagator is not invariant
under longitudinal boosts. From figure \ref{oneplusone_boostinvariant} it
becomes evident as well that the problem does not exhibit the symmetry of
the field. Nevertheless, there are simplifications for the homogeneous 
solution $G_H$. 


\subsection{Ultrarelativistic heavy-ion collisions\label{urhics}}

A standard way to describe an ultrarelativistic heavy-ion collision is to
begin with a current of colour charges moving along the light-cone \cite{kr}. 
In order
to obtain the classical gauge field the Yang-Mills equations have to be
solved in the presence of this current. In Lorenz gauge the field 
consists of two different contributions. 

One of them is made up of the Weizsäcker-Williams sheets. These correspond
to the Coulomb fields of the different colour charges boosted into the
closure of the Lorentz group.
Therewith they become $\delta$-distributions in the light-cone coordinates.
This approximation is justified for the description of particles with low
longitudinal momentum (mid-rapidity). On the one hand they have a low resolution in this
direction. On the other they cannot be comovers of the charges on the
light-cone. For comoving particles the plane wave expansion would not be
applicable \cite{baltz}.

The other contribution is the radiation field in the forward light-cone 
$A^{rad}$. By
virtue of longitudinally boost-invariant boundary conditions on the
light cone, it is sometimes taken as being invariant under longitudinal boosts,
too. However, this is not necessarily the case. But anyway, as has been seen
in section \ref{lbi}, there are no fundamental simplifications for the
propagator. Thus this assumption will not be taken into account in the
following discussion.

In Lorenz gauge, the Weizsäcker-Williams contribution to the gauge-field 
takes the form:

\bmp
\be
A_+^{WW}(x)
&=&
-\frac{g}{2\pi}
\sum_{n_L=1}^{N_L}
t_a(t_a^L)_{n_L}
\delta\left[x_--(b_-^L)_{n_L}\right]
\ln\lambda\left|\vec x_T-(\vec b_T^L)_{n_L}\right|,
\nn
A_-^{WW}(x)
&=&
-\frac{g}{2\pi}
\sum_{n_R=1}^{N_R}
t_a(t_a^R)_{n_R}
\delta\left[x_+-(b_+^R)_{n_R}\right]
\ln{\lambda\left|\vec x_T-(\vec b_T^R)_{n_R}\right|}.
\label{wwsheets}
\ee
\emp

\noindent
There is no contribution to the transverse components. $\lambda$ is an
arbitrary constant, regularising the logarithm which does not appear in 
quantities like the field tensor. The $t_a$ are the generators of $SU(3)_c$. 
The \mbox{$(t_a^{L,R})_n$} represent the colour of the charges.

In general, the presence of a second nucleus leads to a precession of the 
charges of the first nucleus and vice versa which is a manifestation of the
covariant conservation of the current. This leads to further 
sheet-like contributions to the gauge field which can 
be combined with the Weizs\"acker-Williams fields by modifying the charges. 
Up to the next-to-leading order in perturbation theory they read \cite{kr}:

\bmp
\be
(t_a^L)_{n_L}
&=&
(t_a^L)_{n_L}^{in}
-
\alpha_S
\sum_{n_R=1}^{N_R}
f_{abc}(t_b^L)_{n_L}^{in}(t_c^R)_{n_R}^{in}
\theta\left[x_+-(b_+^R)_{n_R}\right]
\ln\lambda|\vec x_T-(\vec b_T^R)_{n_R}|
+
{\cal O}({\alpha_S}^2),
\nn
(t_a^R)_{n_R}
&=&
(t_a^R)_{n_R}^{in}
+
\alpha_S
\sum_{n_L=1}^{N_L}
f_{abc}(t_b^L)_{n_L}^{in}(t_c^R)_{n_R}^{in}
\theta\left[x_--(b_-^L)_{n_L}\right]
\ln\lambda|\vec x_T-(\vec b_T^L)_{n_L}|
+
{\cal O}({\alpha_S}^2).
\nn
\ee
\emp

\noindent
with the initial colours \mbox{$(t_a^{L,R})^{in}_{n_{L,R}}$}.
To all orders, the modifications amount to Wilson lines over the gauge field
along the branches of the light cone. Hence, the precession terms are absent
in adequate, {\it i.e.}, light-cone gauges.
In order to ensure colour neutrality of each nucleus the following relations
have to be satisfied:
\mbox{$
\sum_{n_L=1}^{N_L}
t_a(t_a^L)_{n_L}^{in}
=
0
=
\sum_{n_R=1}^{N_R}
t_a(t_a^R)_{n_R}^{in}
$}.
Frequently, pairs of two charges belonging to the same nucleus are assigned to
each other to form a colour-neutral dipole.
Ultimately, higher-order perturbative calculations for the gauge field as
solution of the Yang-Mills equations are only of value in the presence of
hard energy- or short time-scales. Up to now, non-perturbative solutions for
this problem have been obtained only in transverse lattice calculations
\cite{transverse}. In any case, the general form of the field -- continuous
inside the forward light-cone and singular but integrable on the forward
light-cone -- allows to express the retarded propagator by a finite
number of addends.
Resumming the terms involving the
Weizs\"acker-Williams-like contributions together with the longitudinal
derivatives in the homogeneous part in equation (\ref{retprop}) yields:

\bmp
\be
&&\p\left\{
i\int_{y_0}^{x_0}d\xi_0 \gamma^0
[i\gamma^j\partial_j(x)+\gamma^jk_j
+\ssh A^{WW}(\xi_0,\vec x)+\ssh A^{rad}(\xi_0,\vec x)
-m]
\right\}
=
\nn
&=&
\p\left[
i\int_{y_0}^{x_0}d\xi_0
\p\left\{
i\int_{\xi_0}^{x_0}dz_0 \gamma^0
[i\gamma^3\partial_3(x)+\ssh A^{WW}(z_0,\vec x)]
\right\}\right.
\times
\nn
&&~~~~~~~~~~~~~~~~~~~~\times 
\gamma^0
[i\gamma^J\partial_J(x)+\gamma^jk_j+\ssh A^{rad}(\xi_0,\vec x)-m]
\times
\nn
&&~~~~~~~~~~~~~~~~~\times
\left.
\p\left\{
i\int^{\xi_0}_{x_0}dz_0 \gamma^0
[i\gamma^3\partial_3(x)+\ssh A^{WW}(z_0,\vec x)]
\right\}
\right]
\times
\nn
&&\times
\p\left\{
i\int_{y_0}^{x_0}d\xi_0 \gamma^0
[i\gamma^3\partial_3(x)+\ssh A^{WW}(\xi_0,\vec x)]
\right\}
\label{decompo}
\ee
\emp

\noindent
The path-ordered exponentials containing the Weizs\"acker-Williams
contribution to the vectorpotential can be
written in the form of equation (\ref{mtzero}), because they do not contain
any transverse components. Note that $k_3=0$ has to be used as the
longitudinal momentum has been kept inside the outer path-ordered
exponential. Hence, after replacing $A^{WW}$ by equations (\ref{wwsheets}) and
carrying out the path-ordered integration, one obtains:

\bmp
\be
&&\p\left\{
i\int_{y_0}^{x_0}d\xi_0 \gamma^0
[i\gamma^3\partial_3(x)+\ssh A^{WW}(\xi_0,\vec x)]
\right\}
=
\nn
&=&
\rho^+{\cal P}\prod_{n_L=1}^{N_L}
\exp\left[ia_{n_L}^L(\vec x_T)
\left\{\theta[x_--(b_-^L)_{n_L}]-\theta[y_--(b_-^L)_{n_L}]\right\}\right]
e^{-(x_0-y_0)\partial_3(x)}
+
\nn
&+&
\rho^-{\cal P}\prod_{n_R=1}^{N_R}
\exp\left[ia_{n_R}^R(\vec x_T)
\left\{\theta[x_+-(b_+^R)_{n_R}]-\theta[y_+-(b_+^R)_{n_R}]\right\}\right]
e^{+(x_0-y_0)\partial_3(x)},
\ee
\emp

\noindent
with 
$2\pi a_{n_{L,R}}^{L,R}=-gt_a(t_a^{L,R})_{n_{L,R}}
\ln\lambda\left|\vec x_T-(\vec b_T^{L,R})_{n_{L,R}}\right|$.
It has been assumed that the sums in equation (\ref{wwsheets}) are
ordered in such a way that, if \mbox{$m_L>n_L$} then
\mbox{$(b_-^L)_{m_L}>(b_-^L)_{n_L}$} and likewise if \mbox{$m_R>n_R$} then
\mbox{$(b_+^R)_{m_R}>(b_+^R)_{n_R}$}. The same is to hold for the
products in the previous equation and has been indicated by the ${\cal P}$.
In contrast to equation (\ref{mtzero}), the translation operators are
explicitely shown above. There they had been absorbed in equation
(\ref{mtzeroaux}). 

Making repeated use of the relation
$f[\theta(+x)]=\theta(+x)f(1)+\theta(-x)f(0)$ and subsequent simplifications
lead to:

\bmp
\be
&&\p\left\{
i\int_{y_0}^{x_0}d\xi_0 \gamma^0
[i\gamma^3\partial_3(x)+\ssh A^{WW}(\xi_0,\vec x)]
\right\}
=
\nn
&=&
\rho^+{\cal P}\prod_{n_L=1}^{N_L}
\left[
1
+
\left\{
\theta[x_--(b_-^L)_{n_L}]-\theta[y_--(b_-^L)_{n_L}]
\right\}
\right.
\times
\nn
&&\times
\left.
\left\{
\theta(y_0-x_0)\exp[+ia_{n_L}^L(\vec x_T)]
-
\theta(x_0-y_0)\exp[-ia_{n_L}^L(\vec x_T)]
-
1
\right\}
\right]
e^{-(x_0-y_0)\partial_3(x)}
+
\nn
&+&
\rho^-{\cal P}\prod_{n_R=1}^{N_R}
\left[
1
+
\left\{
\theta[x_+-(b_+^R)_{n_R}]-\theta[y_+-(b_+^R)_{n_R}]
\right\}
\right.
\times
\nn
&&\times
\left.
\left\{
\theta(y_0-x_0)\exp[+ia_{n_R}^R(\vec x_T)]
-
\theta(x_0-y_0)\exp[-ia_{n_R}^R(\vec x_T)]
-
1
\right\}
\right]
e^{+(x_0-y_0)\partial_3(x)}
\ee
\emp

In ultrarelativistic heavy-ion collisions, the colliding nuclei are highly
Lorentz-contracted. Therefore, in the infinite-energy limit
\mbox{$\gamma\rightarrow\infty$}, the "longitudinal 
impact parameters" $(b_\pm^{L,R})_{n_{R,L}}$ go to zero like $\gamma^{-1}$.
However, care has to be taken, as important, because qualitatively different 
contributions to observables could 
be neglected, although the expression for the propagator in this limit would
still be correct.
Taking the limit, the factors involving the differences of
the step functions become the same for all terms of the first and the
second contribution respectively, and thus can be factored out. Afterwards, the
idempotency of the step functions remaining in the product can be used to 
split each contribution into three addends which can be recombined in pairs:

\be
&&\p\left\{
i\int_{y_0}^{x_0}d\xi_0 \gamma^0
[i\gamma^3\partial_3(x)+\ssh A^{WW}(\xi_0,\vec x)]
\right\}
=
\p\left\{-\int_{y_0}^{x_0}d\xi_0\gamma^0\gamma^3\partial_3(x)\right\}
+
\nn
&+&
T^L(x,\pm)[\theta(x_-)-\theta(y_-)]e^{-(x_0-y_0)\partial_3(x)}
+
T^R(x,\pm)[\theta(x_+)-\theta(y_+)]e^{+(x_0-y_0)\partial_3(x)},
\nn
\label{recombi}
\ee

\noindent
with:

\be
T^L(x,\pm)+\rho^+
=
\rho^+
(\pm 1)^{N_L}
{\cal P}\prod_{n_L=1}^{N_L}\exp[\pm ia_{n_L}^L(\vec x_T)]
\nn
T^R(x,\pm)+\rho^-
=
\rho^-
(\pm 1)^{N_R}
{\cal P}\prod_{n_R=1}^{N_R}\exp[\pm ia_{n_R}^R(\vec x_T)],
\ee

\noindent
where $\pm$ in the argument indicates the sign of \mbox{$x_0-y_0$}.
If the two nuclei are modeled by means of
dipoles $N_L$ and $N_R$ are even and the overall factors with the sign
are obsolete.
What is left of the longitudinal structure of the nuclei is the
path-ordering of the products. It would become irrelevant after
averaging over all permutations of the charges in which case it would be
possible to write the product over the exponentials as simple exponential
over the sum of the contributions. 

Putting equation (\ref{recombi}) into equation (\ref{decompo}) allows for
the reconstruction of the homogeneous part of the retarded propagators in the 
radiation field $G^{rad}_H$:

\be
G^{rad}_H(x,y,\vec k)
= 
\p\left\{
i\int_{y_0}^{x_0}d\xi_0 \gamma^0
[i\gamma^j\partial_j(x)+\gamma^jk_j
+\ssh A^{rad}(\xi_0,\vec x)
-m]
\right\}
\ee

\noindent
and the final result reads:

\be
&&G_H(x,y,\vec k)
=
G_H^{rad}(x,y,\vec k)
+
\nn
&&+
\int d^4\xi
G_H^{rad}(x,\xi,\vec k)
[T^L(\xi,\pm)\delta(\xi_-)+T^R(\xi,\pm)\delta(\xi_+)]
G_H^{rad}(\xi,y,\vec k)
+
\nn
&&+
\int d^4\xi d^4\eta
G_H^{rad}(x,\xi,\vec k)
T^L(\xi,\pm)\delta(\xi_-)
G_H^{rad}(\xi,\eta,\vec k)
T^R(\eta,\pm)\delta(\eta_+)
G_H^{rad}(\eta,y,\vec k)
+
\nn
&&+
\int d^4\xi d^4\eta
G_H^{rad}(x,\xi,\vec k)
T^R(\xi,\pm)\delta(\xi_+)
G_H^{rad}(\xi,\eta,\vec k)
T^L(\eta,\pm)\delta(\eta_-)
G_H^{rad}(\eta,y,\vec k).
\ee

\begin{center}
\begin{figure}[h]
\resizebox{!}{6cm}{\includegraphics{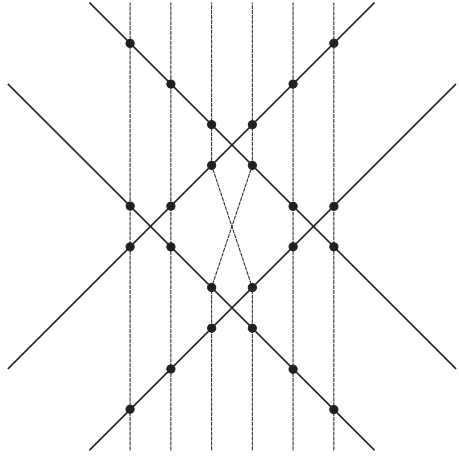}}
\vskip 5mm
\caption{The six contributions to the retarded propagator of highest order in 
the Weizsäcker-Williams fields of two colliding dipoles. The solid lines
depict the trajectories of the charges, the dashed generic integration
paths which, in the forward light-cone, correspond to retarded propagators
in the radiation field.}
\label{pertret}
\end{figure}
\end{center}

\noindent
When the retarded (advanced) propagator is constructed from this homogeneous
part, the minus (plus) sign in the argument of the $T$ has to be taken.
Higher orders in the $T$ cannot contribute, because lines of
constant $x_-$ or $x_+$ can only be crossed once (see the
discussion above). 
In the case of two colliding dipoles taking the ultrarelativistic limit in
order to obtain equation (\ref{recombi}) keeps only integration paths like the 
outer two depicted in figure \ref{pertret}. The other, qualitatively
different paths, in which interactions with left- and right-movers
are intertwined are neglected. With finite 
"longitudinal impact parameters",
every charge would have to be treated on its own and the previous equation 
would contain contributions with up to \mbox{$N_L+N_R$} insertions. 

For a given field, the retarded propagator in the radiation field 
could now be obtained with the methods presented 
earlier in this section.
The exact solution for the radiation field would be needed. Finally
averaging over all configurations (current and hence field) would yield the
desired observables. However, the transverse lattice calculation required at
that point is beyond the scope of the present paper.


\section{Fermion-antifermion pair production}

In this section, the findings for the full propagator in a space-time
dependent field are interpreted in view of the problem of particle
production by vacuum polarisation. Such a calculation can be understood
in two different ways. Either the field is an external field in the strict
sense, or the field is the self-consistent solution of a set of equations.
In the first case it is governed entirely by the characteristics of the
physical system without the phenomenon of particle creation. Subsequently it
can be used to determine the spectrum of produced particles. An approach of
this kind is justified if particle production represents only a perturbation
which must be checked eventually. In the second case, given the solution for
the set of equations in form of the classical field is known, the question
is how many particles were created in the process. Here, the set of equations
would include the Yang-Mills equations with an initially present current and
the current induced by the fermions and antifermions.
In this section the first way is to be pursued, because the back-reaction in
the investigated system is suppressed by higher powers of the coupling
constant not accompanied by powers of the vector potential.

According to \cite{derivation} the expectation value for the number of
produced pairs can be determined by starting out with a negative energy
fermion with four-momentum $q$ corresponding to a positive energy 
antifermion:

\be
\psi_q(y)=v(q)e^{-iq\cdot y},
\label{psiinitial}
\ee

\noindent
with the unit spinor $v(q)$. After its propagation through the field, calculate
its overlap with a positive energy fermion solution with four-momentum 
$p$:

\be
\psi_p(x)=u(p)e^{+ip\cdot x},
\label{psifinal}
\ee

\noindent
with the unit spinor $u(p)$. Wave-function solutions of the Dirac equation at 
$x$ and $y$ are connected by the retarded propagator according to:

\be
\psi(x)=\int d^3y G_R(x,y)\gamma^0\psi_q(y).
\ee

\noindent
Hence, the required overlap reads:

\be
M(p,q)=\int d^3x d^3y \psi^\dagger_p(x) G_R(x,y) \gamma^0 \psi_q(y).
\label{overlap}
\ee

With the definition of the retarded one-particle scattering-operator 
${\cal T}_R$ 

\be
G_R(x,y)
=
G_R^0(x-y)
+
\int d^4\xi d^4\eta G_R^0(x-\xi){\cal T}_R(\xi,\eta) G_R^0(\eta-y),
\label{ropso}
\ee

\noindent
the overlap can be rewritten as:

\be
M(p,q)
~=~
\int d^4\xi d^4\eta 
\bar\psi_p(\xi) {\cal T}_R(\xi,\eta) \psi_q(\eta)
~=~
\bar u(p){\cal T}_R(p,-q)v(q),
\ee

\noindent
because $\psi_q$ and $\psi_p$ are orthogonal to each other and 
are solutions of the free Dirac equation, whence:

\be
\int d^3y G_R^0(x-y)\gamma^0\psi_q(y)&=&\psi_q(x)
\nn
\int d^3y \bar\psi_p(y)\gamma^0G_R^0(y-x)&=&\bar\psi_p(x).
\ee

\noindent
In the last step the definitions of 
$\psi_q$ and $\psi_p$ in equations (\ref{psiinitial}) and
(\ref{psifinal}) had been inserted and the integrations carried out. This led
to the Fourier transformation of the single-particle scattering operator. 
The square of this overlap $|M(p,q)|^2$ summed/averaged over all colours and
spins is equal to the differential expectation value for the number of produced pairs
where the particle carries the momentum $p$ and the antiparticle the
momentum $q$. After integration over the phase space of the particle 
(antiparticle) one obtains the spectrum of produced antiparticles
(particles). Integration over all momenta finally gives
the expectation value for the number of produced pairs:

\be
\left<n\right>
=
\int 
\frac{d^3q}{2(2\pi)^3\omega_q} 
\frac{d^3p}{2(2\pi)^3\omega_p}
\left|\bar{u}(q){{\cal T}}_R(q,-p)v(p)\right|^2,
\label{expectation}
\ee

\noindent
where $\omega_p=\sqrt{|\vec p|^2+m^2}$ and $\omega_q=\sqrt{|\vec q|^2+m^2}$.

~\\

In order to address the issue of gauge invariance, rewrite the previous
expression as \cite{baltz}:

\be
\left<n\right>
=
-\int\frac{d^3q}{(2\pi)^32\omega_q}d^4xd^4ye^{iq\cdot(x-y)}
{\rm tr}\left\{
(\ssh q+m)[i\ssh\partial(x)-m]G_{+-}(x,y)[i\ssh\partial(y)-m]\right\},
\nn
\label{reduction}
\ee

\noindent
where the derivatives are only acting on the propagator. 
The only coloured
and hence gauge-dependent part of equation (\ref{reduction}) is the
propagator, even after taking the trace over its colour indices:

\be
{\rm tr}\left\{G(x,y)\right\}
\rightarrow
{\rm tr}\left\{G'(x,y)\right\}
=
{\rm tr}\left\{\Omega(x)G(x,y)\Omega^\dagger(y)\right\}
\neq
{\rm tr}\left\{G(x,y)\right\}
\label{gaugetrafo}
\ee


\subsubsection{Gauge-dependent states / Wilson links\label{wilson}}

Sometimes, the gauge invariance of the previous expressions is ensured by 
replacing
the gauge-independent free states in equations (\ref{psiinitial}) and
(\ref{psifinal}) by gauge dependent states $\tilde\psi$, especially fermion 
and antifermion
solutions in pure gauge fields, {\it i.e.}, vacuum configurations of the
gauge field: 

\be
\left\{
i\ssh\partial(x)+\Omega(x)[\ssh\partial(x)\Omega^\dagger(x)]-m
\right\}\tilde\psi(x)
=0.
\ee

\noindent
The gauge transformations associated with them cancel those coming from the
propagator in equation (\ref{overlap}) exactly:

\be
\tilde M(p,q)
&=&
\int d^3xd^3y \tilde\psi^\dagger_p(x)G_R(x,y)\gamma^0\tilde\psi_q(y)
\rightarrow
\int d^3xd^3y \tilde\psi^{\dagger\prime}_p(x)G_R^\prime(x,y)\gamma^0
\tilde\psi_q^\prime(y)
=
\nn
&=&
\int d^3xd^3y
\tilde\psi^\dagger_p(x)\Omega^\dagger(x)\Omega(x)G_R(x,y)\Omega^\dagger(y)
\gamma^0\Omega(y)\tilde\psi_q(y)
=
\tilde M(p,q)
\ee

A priori, the gauge in which \mbox{$\tilde\psi(x)$} coincides with a free
solution can be chosen arbitrarily, whence it must be specified in
order to allow for an interpretation, which therefore is not unique. This
goes hand-in-hand with the loss of an interpretation as states with a
definite number of partons. In all but the one gauge of reference an
arbitrary number of additional gluons is included in them. Further, when
deriving equation (\ref{expectation}) $q$ and $p$ were the momenta of the
fermion and the antifermion, respectively. Here, this correspondence remains
obvious only in one gauge, which thus would be singled out.

This method is equivalent to including Wilson lines with the
correlators. For an adequate choice of these gauge links -- especially their
end-points -- the gauge
transformations connected to them cancel with those of the two-point
functions. There the supplementary degree of freedom is the path along which
it is evaluated.


\subsubsection{Average over the gauge group\label{gaugegroup}}

A way to guarantee a gauge invariant-expression without sacrificing the
requirement of purely fermionic states is given by averaging the expression
(\ref{reduction}) for the expectation value over all gauges,
because every supplementary gauge transformation $\omega(x)$ can then be
absorbed by shifting the functional-integral measure:

\be
\int [d\Omega]{\rm tr}\left\{\omega(x)G(x,y)\omega^\dagger(y)\right\}
=
\int [d(\omega\Omega)]{\rm tr}\left\{G(x,y)\right\}
=
\int [d\Omega']{\rm tr}\left\{G(x,y)\right\}.
\label{gaugeaverage}
\ee

As a first step, carry out a coordinate transformation from $x$ and $y$ to
\mbox{$\xi=x-y$} and \mbox{$\zeta=(x+y)/2$}. Subsequently look at 
$d\left<n\right>/d^4\xi$.
Any SU(N) gauge transformation can be parametrised according to

\be
\Omega(x)
=
\exp\{-it^a\theta^a(x)\}
=
\sum_{n=1}^N\left|n(x)\right>\left<n(x)\right|\exp\{-i\lambda_n(x)\}
\ee

\noindent
with $t^a$ the generators of the gauge algebra and
where $\lambda_n(x)$ are real eigenvalues and $\left|n(x)\right>$
orthonormalised eigenvectors of the matrix $t^a\theta^a(x)$\footnote{These 
eigenvalues and eigenvectors are not the same as those in
section \ref{wilsonexample}.}. 
Functional integration of $d\left<n\right>/d^4\xi$
over the eigenvalues in equation (\ref{gaugetrafo})  
yields a non-zero result for $x=y$:

\be
\left.
\frac{d\left<\bar{n}\right>}{d^4\xi}
\right|_{\xi=0}
&=&
-\int\frac{d^3q}{(2\pi)^32\omega_q}
d^4\zeta
\times
\nn
&&\times
{\rm tr}\left\{
(\ssh q+m)
\left[i\left.\ssh\partial(\zeta_1)\right|_{\zeta_1=\zeta}-m\right]
G_{+-}(\zeta_1,\zeta_2)
\left[i\left.\ssh\partial(\zeta_2)\right|_{\zeta_2=\zeta}-m\right]
\right\}
\nn
\ee

\noindent
Already in its present form the last expression is gauge invariant.
Averaging over the eigenvectors $\left|n(x)\right>$ is not needed, because
the trace already takes care of colour neutrality. Now the two gauge
transformations in equation (\ref{gaugetrafo}) are evaluated at the same
point and drop out of the trace. 

This result resembles calculations on the lattice. There the results contain
the four-dimensional lattice-volume ${\cal V}$ as an overall factor. Hence,
in that case one would look at a quantity like $\left<\bar{n}\right>/{\cal
V}$.

With the integrand in momentum space, the previous result becomes:

\be
\left.
\frac{d\left<\bar{n}\right>}{d^4\xi}
\right|_{\xi=0}
&=&
-\int\frac{d^3q}{(2\pi)^32\omega_q}
\frac{d^4l}{(2\pi)^4}
{\rm tr}\left\{
(\ssh q+m)
{G_R^0}^{-1}(l)
G_{+-}(l,l)
{G_A^0}^{-1}(l)
\right\}
\nn
\ee

\noindent
instead of 

\be
\left<n\right>
=
\int\frac{d^3q}{(2\pi)^32\omega_q}
{\rm tr}\left\{(\ssh q+m) {G_R^0}^{-1}(q) G_{+-}(q,q) {G_A^0}^{-1}(q)
\right\},
\ee

\noindent
which would be obtained directly from equation (\ref{reduction}).

Making use of equation (\ref{trafo}) 
(\mbox{$\sigma_1=+$}, \mbox{$\sigma_2=-$}) as well as
\cite{baltz}\footnote{The advanced one-particle scattering-operator is
defined in analogy with the retarded in equation (\ref{ropso}).}:

\be
{\cal T}_A(q,p)-{\cal T}_R(q,p)
=
\int\frac{d^4k}{(2\pi)^3}\delta(k^2-m^2){\rm sgn}(k_0)
{\cal T}_R(q,k)(\ssh k+m){\cal T}_A(k,p)
\ee

\noindent
and:

\be
G_S(q,p)
=
\int\frac{d^4k}{(2\pi)^3}\delta(k^2-m^2)
G^0_R(q){\cal T}_R(q,k)(\ssh k+m){\cal T}_A(k,p)G^0_A(p), 
\ee

\noindent
gives:

\be
\left.\frac{d\left<\bar{n}\right>}{d^4\xi}\right|_{\xi=0}
=
\int
\frac{d^4q}{(2\pi)^4}\theta(+q_0)2\pi\delta(q^2-m^2)
\frac{d^4k}{(2\pi)^4}\theta(-k_0)2\pi\delta(k^2-m^2)
\frac{d^4l}{(2\pi)^4}
\times~~~~~~~~~~
\nn
\times
{\rm tr}\{(\ssh q+m){\cal T}_R(l,k)(\ssh k+m){\cal T}_A(k,l)\}.
\label{averaged}
\ee

\noindent
This equation can be interpreted as the fermionic current induced by the
external field from which the on-shell component \mbox{($q^2=m^2$)} is 
projected out with the extra condition that the current's momentum is 
on the mass shell 
once \mbox{($k^2=m^2$)}. Opposed to that, in equation (\ref{expectation})
the momentum is forced to be on-shell twice. 

Under a local gauge transformation $\Omega(x)$, the full propagator
transforms homogeneously [see equation (\ref{gaugetrafo})].  The
considerations concerning gauge invariance remain unchanged, if this
behaviour is preserved in an approximation. In the case where, in a given
gauge and/or Lorentz frame a certain expansion should be exact, this
requirement is trivially satisfied. In general, if the expansion is based on
powers of the covariant derivatives and/or the mass, covariance is always 
ensured.  This is the case for equation (\ref{mtzero}), for expansions of
equation (\ref{smalltransversemass}) if the transverse momentum is omitted,
and for approximations like equation (\ref{firstorder}) if the transverse
derivatives are included with \mbox{${\cal M}_T$} [after adapting equations
(\ref{mtzeroaux}) and (\ref{untranslated})]. If the approximate propagator
should not transform homogeneously, still a gauge invariant result can be
obtained by averaging over all gauges. However, in that case the situation
has to be reinvestigated, because the condition $x=y$ need not
be sufficient to guarantee gauge invariance there.


\subsection{An example}

The radiation field in an ultrarelativistic heavy-ion collision has to be
determined in transverse lattice calculations which are beyond the scope of
the present paper. Hence, the previously determined expressions for
the propagator and the quantities derived from it are to be evaluated for a 
model for the radiation field.
Along the lines of the discussion in section \ref{urhics} a longitudinally
boost-invariant field is chosen:

\be
A_\pm=\pm a \frac{x_\pm}{\tau_0} e^{-x_+x_-/{\tau_0}^2}.
\label{gaugefield}
\ee

\noindent
The Lorenz $\partial\cdot A=0$ and the Fock-Schwinger $x\cdot A=0$ 
gauge-conditions are satisfied by this field.
In its presence the one-particle scattering-operator defined in equation
(\ref{ropso}) with the propagator according to the lowest order expansion in
the transverse components in section \ref{urhics} reads:

\be
{\cal T}_R(p,q)
=
(2\pi)^2\delta^{(2)}(\vec p_T-\vec q_T){\cal T}_R(\tilde p,\tilde q)
\label{sopso} 
\ee

\noindent
with:

\be
\sqrt{2}{\cal T}_R(\tilde p,\tilde q)
=
\gamma_+{\cal T}_R^+(\tilde p,\tilde q)
+
\gamma_-{\cal T}_R^-(\tilde p,\tilde q)
\ee

\noindent
where:

\be
{\cal T}_R^\pm(\tilde p,-\tilde q)
&=&
\mp a\tau_0\frac{
f(r_+r_-{\tau_0}^2)
}{
(r_+r_-{\tau_0}^2)
r_\pm
}
-
\nn
&&-
a^2
\sum_{\mu\nu}
\frac{(\pm ia\tau_0)^\mu}{\mu!}
\frac{(\mp ia\tau_0)^\nu}{\nu!}
\frac{
f[r_+r_-{\tau_0}^2/(\mu+\nu+2)]
-
f[p_\pm r_\mp{\tau_0}^2/(\mu+\nu+2)]
}{
{r_\pm}^2
[(\mu+1)q_\mp-(\nu+1)p_\mp]
},
\label{copso}
\nn
\ee

\noindent
with
\mbox{$
r_\pm=p_\pm+q_\pm
$}.
In equation (\ref{sopso}) the omnipresent Dirac-distribution, originating
from the transverse homogeneity of the field, is factored out. The
expressions for the two components of the one-particle scattering-operator
can be resummed partially into hypergeometric functions. $f$ involves
the exponential-integral function ${\rm Ei}$:
\mbox{$
f(\sigma)
=
\sigma^2
e^{-\sigma}
{\rm Ei}(\sigma)
-
\sigma.
$}
In accordance with the discussion in the previous section, the approximation
to the retarded propagator used here transforms like the
full solutions, {\it i.e.}, homogeneously. Higher order contributions to the retarded propagator 
$G_R(x,y)$ are also available analytically. They are qualitatively 
similar to the lowest order investigated here. However, the terms belonging
to the different orders cannot be resummed easily. Further, a related
approach to the problem of particle production investigated in \cite{field} 
turned out to be in good agreement with the full solution.


\subsubsection{Gauge-dependent states / Wilson links\label{wilsonexample}}

An interpretation of equations (\ref{sopso}-\ref{copso}) put into the
integrand of equation (\ref{expectation}):

\be
\left|\bar{u}(q){\cal T}_R(q,-p)v(p)\right|^2
=
\sum_n
\left[
8(p\cdot{\cal T}_R)(p\cdot {{\cal T}_R]}^\dagger)
-
4(p\cdot q-m^2)({\cal T}_R\cdot{{\cal T}_R}^\dagger)]
\right]
\ee

\noindent
along the lines of the discussion in section \ref{wilson}
would be possible if the gauge of reference was chosen to be the Lorenz or
the Fock-Schwinger gauge. Here, in these gauges, the vacuum configuration of 
the gauge field vanishes identically per definition, whence the fermion 
solutions $\tilde\psi$ in the vacuum configuration coincide with the free 
solution. The last expression is summed over the eigenvalues $\lambda_n$ of 
the colour matrix \mbox{$
a=\sum_n\lambda_n\left|n\right>\left<n\right|
$}\footnote{These eigenvalues and eigenvectors are not the same as those in
section \ref{gaugegroup}.}. 

In the case, where a light-cone gauge, say $A_+=0$ was taken, one would have
to resort to the expression of the one-particle scattering-operator in that
gauge:

${\cal T}_R^-=0$

\noindent
and

\bmp
\be
{\cal T}_R^+(\tilde p,-\tilde q)
&=&
-2a\tau_0\frac{
f(r_+r_-{\tau_0}^2)
}{
(r_+r_-{\tau_0}^2)
r_+
}
-
\nn
&&-
4a^2
\sum_{\mu\nu}
\frac{(+2ia\tau_0)^\mu}{\mu!}
\frac{(-2ia\tau_0)^\nu}{\nu!}
\frac{
f[r_+r_-{\tau_0}^2/(\mu+\nu+2)]
-
f[p_+r_-{\tau_0}^2/(\mu+\nu+2)]
}{
{r_+}^2
[(\mu+1)q_--(\nu+1)p_-]
}.
\label{lc}
\nn
\ee
\emp

\noindent
In other words, the Wilson line taken along the $x_-$-direction becomes
unity.


\subsubsection{Average over the gauge group}

For a calculation according to section \ref{gaugegroup} one can start in any
gauge.
In momentum space, the advanced single-particle scattering-operator is
linked to the retarded via \mbox{${\cal T}_A(p,q)=-{\cal T}_R^*(q,p)$}.
Starting from equation (\ref{lc})
the integration over the momentum $l$ and the trace can be executed to
yield:

\bmp
\be
&&\frac{2}{8q_+k_+}
\int\frac{d^2\tilde l}{(2\pi)^2}
{\rm tr}\{(\ssh q+m){\cal T}_R(\tilde l,\tilde k)(\ssh k+m)
{\cal T}_A(\tilde k,\tilde l)\}
=
\nn
&=&
(a\eta)^2
+
\sum_{\mu\nu}\frac{(+2ia\tau_0)^\mu}{\mu!}\frac{(-2ia\tau_0)^\nu}{\nu!}
\times
\nn
&&\times
\left(
+
4i(a\tau_0)^3
\left\{
\frac{\eta^2(\mu+2)+2i{\tau_0}^2k_-\eta}{2{\tau_0}^2(\mu+2)^2(\mu+\nu+3)}
-
{\tau_0}^2{k_-}^2\ln\left[1+i\frac{\eta(\mu+2)}{k_-{\tau_0}^2}\right]
\right\}
-
\right.
\nn
&&~~~~-
4i(a\tau_0)^3
\left\{
\frac{\eta^2(\nu+2)+2i{\tau_0}^2k_-\eta}{2{\tau_0}^2(\nu+2)^2(\mu+\nu+3)}
-
{\tau_0}^2{k_-}^2\ln\left[1+i\frac{\eta(\nu+2)}{k_-{\tau_0}^2}\right]
\right\}
+
\nn
&&~~~~+
\frac{16(a\tau_0)^4}{\mu+\nu+4}
\left\{
\frac{\eta^2(\nu+\mu+6)}{4{\tau_0}^2(\nu+3)(\mu+3)}
+
\frac{4ik_-\eta(\nu+\mu+6)(\nu-\mu)}{2(\nu+3)^2(\mu+3)^2}
-
\right.
\nn
&&~~~~~~~~~~~~
\left.
\left.
-
\frac{{\tau_0}^2{k_-}^2}{(\nu+3)^3}
\ln\left[1-i\frac{\eta(\nu+3)}{k_-{\tau_0}^2}\right]
-
\frac{{\tau_0}^2{k_-}^2}{(\mu+3)^3}
\ln\left[1+i\frac{\eta(\mu+3)}{k_-{\tau_0}^2}\right]
\right\}
\right).
\nn
\ee
\emp

\noindent
The quantity $\eta$ regularises the expression. The divergencies are due to
the symmetry (boost invariance) of the field and to the expansion of the
propagator in the transverse mass.
It can be checked -- {\it e.g.} by interchanging the indices $\nu$ and $\mu$ -- that
the last expression is in fact real, as it should be. Further the sums
correspond to hypergeometric functions and their special cases like
gamma functions. The logarithms of complex argument are linked to logarithms 
of the modulus of the argument or arcus tangents  
respectively. However the above is still the most compact representation.
The final result has to be summed over the eigenvalues of the
colour matrix $a$ as before.

The momentum $k$ cannot be identified with the antiparticle momentum $p$.
This can be seen after integrating over the plus and minus components of the
momenta $k$ and  $q$ and taking the resulting expression at vanishing
transverse mass \mbox{$\sqrt{|\vec k_T|^2+m^2}=0$}:

\bmp
\be
&&\frac{(2\pi)^4{\tau_0}^2}{{L_T}^2\eta^2\kappa^2}
\left.
\frac{d\left<\bar{n}\right>}{d^4\xi d^2k_T d^2q_T}
\right|_{\xi=0=\sqrt{|\vec k_T|^2+m^2}}
=
(a\tau_0)^2
+
\sum_{\mu\nu}\frac{(+2ia\tau_0)^\mu}{\mu!}\frac{(-2ia\tau_0)^\nu}{\nu!}
\times
\nn
&&~~~~~~~~~~~~~~~~~~~~~~~~~~~~~~~\times
\left(
\frac{2i(a\tau_0)^3(\nu-\mu)}{(\mu+2)(\nu+2)(\mu+\nu+3)}
+
\frac{4(a\tau_0)^4(\nu+\mu+6)}{(\nu+3)(\mu+3)(\mu+\nu+4)}
\right).
\nn
\ee
\emp

\noindent
Use has been made of the relation 
\mbox{$(2\pi)^2[\delta^{(2)}(\vec k_T-\vec l_T)]^2
=
{L_T}^2\delta^{(2)}(\vec k_T-\vec l_T)$} with the transverse length $L_T$.
$\kappa$ serves to regularise the integrals. In the final result $\eta$ and
$\kappa$ occur exclusively in the dimensionless product $\eta\kappa$.
Due to Noether's theorem, in the absence of transverse components and 
variations of the field, the particle and antiparticle should be produced in 
strict back-to-back configurations only. Hence, the right-hand side of the
last equation should have support merely for \mbox{$\vec q_T=-\vec k_T=0$},
but this is not the case. In fact it does not depend on $q_T$ at all.
Already in the general expression (\ref{averaged}), the dependence on
the particle momentum $q$ is only via the phase-space volume and the factor
in the trace. Therefore, this quantity can only be interpreted consistently
after integration over the momentum variable $k$. For the present example, all the
integrals could be carried out analytical. Here the result for massless
fermions in a system with one temporal and one spatial (longitudinal) 
dimension is spelled out, where it is the exact result:

\be
&&\frac{(2\pi)^2{\tau_0}^2}{\eta^2\kappa^2}
\left.
\frac{d\left<\bar{n}\right>}{d^2\tilde\xi}
\right|_{\xi=0}^{(1+1)}
=
4(a\tau_0)^2
+
\frac{3}{2}[1-\cos(2a\tau_0)]
\ee

For strong fields and/or slow decay-times, the gauge-field (\ref{gaugefield})
corresponds to an (almost) constant electrical field. In that limit the last
expression is dominated by the quadratic term. This observation is in 
correspondance with the
behaviour of the constant-field result by Schwinger \cite{schwinger}
in the case of vanishing transverse mass.


\section{Summary}

The creation of fermion-antifermion pairs due to vacuum polarisation in the
presence of classical fields has been investigated starting from the
advanced and retarded fermion-propagators in the field. The needed
propagators were derived as Green's function solutions of the Dirac equation
to all orders in the arbitrarily space-time dependent field. The required
propagators were chosen by imposing the respective boundary conditions. In the
course of the calculations, $U(N)$-generators have been admitted for the 
charge-space of the gauge field. Thus QED and QCD effects can be taken into
account simultaneously. In terms of creation processes electron-positron
production as well as quark-antiquark chromo- and photo-production can be
adressed. After a discussion of the general features of the propagator and
its constituents, like the generalised translation operator, its
characteristics have been treated in great detail. In order to gain further
insight into the structure of the obtained Green's functions and in order to
get the link to the results for purely time-dependent fields in
\cite{field}, various expansion schemes have been derived from the full
expression. Namely, these are the weak-field, the strong field, the
gradient, and the Abelian approach as well as some combinations thereof.
Further, also for the same reasons as above, the features of the propagator 
are investigated thoroughly in different classes of field configurations.
This includes situations that are dominated by temporal and longitudinal
degrees of freedom and last but not least longitudinal boost-invariance of
the system. This feature is frequently of importance for the description of 
ultrarelativistic heavy-ion collisions in the semiclassical limit. Therefore
this case has been spread out in detail in a seperate subsection. However,
it has been found that the assumption of longitudinal boost-invariance of
the gauge field does not lead to significant simplifications for the
propagator. The reason herefore is that its boundary condition does not
display this symmetry. Contrary to that there are simplifications for the
homogeneous solutions.

The issue of the gauge invariance of the expression for the expectation
value for the number of fermion-antifermion pairs produced in the presence
of the field is discussed in detail. These expressions are functionals of
the propagators which have been derived above. These two-point functions are
the gauge dependent (coloured) objects. Even the trace over them is in
general not gauge invariant. Hence, two different ways to generate
gauge-invariant quantities are discussed. On the one hand, this can be
achieved by projecting onto gauge-dependent states instead of free fermion
solutions. This method is equivalent to the inclusion of Wilson lines. In a
given gauge of reference the new expression coincides with the bare one.
This is where it is evaluated. Nevertheless, the interpretation as purely
fermionic states is lost, because in all but the gauge of reference, in the
assymptotic states gluons
are included with the fermions. On the other hand, in order to preserve
purely fermionic states, one can average the expectation value over the
gauge group. In both approaches, it seems not to be possible to identify the
particle- and antiparticle-momentum simultaneously. While the momentum of
the particle can be identified, in the second case, the supplementary momenta, 
do not satisfy Noether's theorem. In the first case, the additional gluons
turn the momentum-like variable more into a continuous parameter for the
numbering the states than into the antiparticle momentum.

Finally, the various expressions are evaluated in the presence of a model
for the radiation field in an ultrarelativistic heavy-ion collision, because
the exact field configuration, for example in the McLerran-Venugopalan model is
only availabe numerically. For said field, an analytic expression for the
one-particle scattering operator to lowest order in an expansion in the
transverse components of the covariant derivatives and the mass is given. This 
expansion
preserves the behaviour under gauge transformations of the considered
quantities. Higher order terms for the propagator in this field are also
available analytically. In the framework involving the gauge-dependent states
(Wilson lines), after carrying out the traces over the colour and the
Clifford algebra, this result provides a two-parameter inclusive distribution
of the states as explained before. For the method, where the average over the 
gauge group has been taken, the integration over all momentum variables can
be carried out analytically. In a certain limit, the present model field-tensor 
becomes constant. There, a behaviour quadratic in the field
strenght is obtained. For vanishing transverse mass, the same holds for the 
constant-field Schwinger-formula.


\section*{Acknowledgements}

The author would like to thank A.~Mosteghanemi for his help.
Informative discussions with F.~Guerin, G.~Korchemsky, A.~Mishra, 
A.~Mueller, J.~Reinhardt, D.~Rischke, D.~Schiff, and S.~Schramm are 
acknowledged gratefully. This 
work has been supported financially by the DAAD (German Academic Exchange 
Service).



\appendix

\section*{Figures}

\newpage

\newpage

\end{document}